\documentclass[aps,prl,reprint,superscriptaddress,amsmath,amssymb,longbibliography]{revtex4-2}
\usepackage{graphicx,bm,mathrsfs,hyperref,tikz,scalerel}
\hypersetup{colorlinks=true, citecolor=blue, urlcolor=blue, linkcolor=blue} 
\renewcommand{\section}[1]{{\par\it #1.---}\ignorespaces}
\usetikzlibrary{svg.path}
\definecolor{orcidlogocol}{HTML}{A6CE39}
\tikzset{orcidlogo/.pic={
		\fill[orcidlogocol] svg{M256,128c0,70.7-57.3,128-128,128C57.3,256,0,198.7,0,128C0,57.3,57.3,0,128,0C198.7,0,256,57.3,256,128z};
		\fill[white] svg{M86.3,186.2H70.9V79.1h15.4v48.4V186.2z}
		svg{M108.9,79.1h41.6c39.6,0,57,28.3,57,53.6c0,27.5-21.5,53.6-56.8,53.6h-41.8V79.1z M124.3,172.4h24.5c34.9,0,42.9-26.5,42.9-39.7c0-21.5-13.7-39.7-43.7-39.7h-23.7V172.4z}
		svg{M88.7,56.8c0,5.5-4.5,10.1-10.1,10.1c-5.6,0-10.1-4.6-10.1-10.1c0-5.6,4.5-10.1,10.1-10.1C84.2,46.7,88.7,51.3,88.7,56.8z};}}
\newcommand\orcid[1]{\href{https://orcid.org/#1}{\mbox{\scalerel*{\begin{tikzpicture}[yscale=-1,transform shape]\pic{orcidlogo};\end{tikzpicture}}{|}}}}

\begin{document}
\title{Self-Discharging Mitigated Quantum Battery}
\author{Wan-Lu Song\orcid{0000-0002-6437-9748}}
\affiliation{Department of Physics, and Key Laboratory of Intelligent Sensing System and Security (Ministry of Education), Hubei University, Wuhan 430062, China}
\author{Ji-Ling Wang}
\affiliation{Department of Physics, and Key Laboratory of Intelligent Sensing System and Security (Ministry of Education), Hubei University, Wuhan 430062, China}
\author{Bin Zhou\orcid{0000-0002-4808-0439}}
\email{binzhou@hubu.edu.cn}
\affiliation{Department of Physics, and Key Laboratory of Intelligent Sensing System and Security (Ministry of Education), Hubei University, Wuhan 430062, China}
\affiliation{Wuhan Institute of Quantum Technology, Wuhan 430206, China}
\author{Wan-Li Yang}
\email{ywl@wipm.ac.cn}
\affiliation{State Key Laboratory of Magnetic Resonance and Atomic and Molecular Physics, Innovation Academy for Precision Measurement Science and Technology, Chinese Academy of Sciences, Wuhan 430071, China}
\author{Jun-Hong An\orcid{0000-0002-3475-0729}}
\email{anjhong@lzu.edu.cn}
\affiliation{Key Laboratory of Quantum Theory and Applications of MoE, Lanzhou Center for Theoretical Physics, and Key Laboratory of Theoretical Physics of Gansu Province, Lanzhou University, Lanzhou 730000, China}

\begin{abstract}
   As a quantum thermodynamic device that utilizes quantum systems for energy storage and delivery, the quantum battery (QB) is expected to offer revolutionary advantages in terms of increasing the charging power and the extractable work by using quantum resources. However, the ubiquitous decoherence in the microscopic world inevitably forces the QB to spontaneously lose its stored energy. This is called the self-discharging of the QB and severely limits its realization. We propose a QB scheme based on the nitrogen-vacancy center in diamond, where the electronic spin serves as the QB. Inspired by our finding that the coherent ergotropy decays more slowly than the incoherent ergotropy, we reveal a mechanism to enhance the inherent robustness of the QB to the self-discharging by improving the ratio of coherent ergotropy to total ergotropy. The unique hyperfine interaction between the electron and the native $^{14}$N nucleus in our scheme allows one to coherently optimize this ratio. Mitigating the self-discharging and optimizing the extractable work simultaneously, our results pave the way for the practical realization of the QB.
\end{abstract}

\maketitle

\section{Introduction}
Quantum thermodynamics integrates the fascinating laws from quantum mechanics with the principles of work and energy from thermodynamics \cite{PhysRevLett.115.210403,Bera2017,Alicki2018}. Quantum effects make microscopic systems exhibit thermodynamic behaviors that are distinctly different from or surpass those of macroscopic systems \cite{PhysRevLett.112.030602,PhysRevLett.122.110601,Micadei2019,Mayer2023}. They aid in the design of novel devices at the atomic scale \cite{Alicki_1979,PhysRevLett.105.130401,doi:10.1126/science.aad6320,Maslennikov2019} and hold immeasurable value for the development of information technology, energy science, and other future technologies \cite{Parrondo2015,Millen_2016,PRXQuantum.3.020101}. 

As a typical quantum thermodynamic device designed to store and supply energy, the QB has attracted wide attention \cite{PhysRevLett.111.240401,Guryanova2016,PhysRevLett.118.150601,Campaioli2018,PhysRevLett.122.210601,PhysRevLett.124.130601,PhysRevLett.125.040601,PhysRevLett.125.236402,PhysRevLett.127.100601,PhysRevLett.131.030402,PhysRevLett.132.210402,PhysRevLett.132.240401,RevModPhys.96.031001,PhysRevLett.133.243602,PhysRevLett.134.180401}. The utilization of quantum resources \cite{RevModPhys.81.865,RevModPhys.89.041003,RevModPhys.91.025001} has been put forward as a potential way to improve the charging power and extractable work of batteries \cite{PhysRevA.101.032115,PhysRevE.102.052109,PhysRevB.102.245407,PhysRevE.104.024129,PhysRevA.104.032207,PhysRevB.104.245418,PhysRevB.105.115405,PhysRevA.106.062609,PhysRevA.107.022215,PhysRevA.108.052213,PhysRevA.110.022433,PhysRevLett.133.197001,PhysRevLett.134.130401}, which could bring a revolutionary upgrade to modern energy devices. It has been revealed that entangling charging operations can accelerate the charging process to achieve superlinear power scaling in the number of the QBs \cite{PhysRevLett.120.117702,PhysRevLett.128.140501}. Entangling unitary controls can extract more work than individual ones \cite{PhysRevE.87.042123}. However, the environment-induced decoherence inevitably causes the spontaneous loss of the QB energy, which is called the self-discharging of the QB \cite{PhysRevE.103.042118,PhysRevE.105.054115,PhysRevE.109.054132}. Several schemes, e.g., Floquet engineering \cite{PhysRevA.102.060201}, quantum reservoir engineering \cite{PhysRevLett.132.090401}, feedback control \cite{PhysRevE.106.014138}, and dark state \cite{PhysRevApplied.14.024092}, have been proposed to suppress the impact of decoherence on the QB but require auxiliary systems to act as chargers. However, the charger-battery entanglement is the main limiting factor for the task of work extraction \cite{PhysRevLett.122.047702}. Therefore, in order to promote the physical realization of the QB \cite{doi:10.1126/sciadv.abk3160,batteries8050043,PhysRevA.106.042601,PhysRevLett.131.240401,PhysRevLett.131.260401,PhysRevA.109.062614}, it is a key issue to enhance the inherent robustness of the QB to the self-discharging in a direct charging protocol.

\begin{figure}[t]
    \centering
    \includegraphics[width=0.95\linewidth]{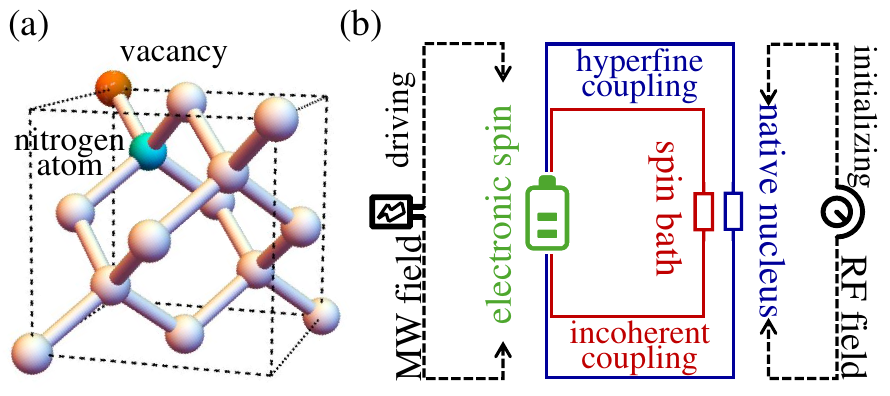}
    \caption{(a) Local view of the NV center in diamond, which contains an electronic spin with $S=1$ and a native $^{14}$N nuclear spin with $I=1$. (b) Taking the electronic spin as the QB, the incoherent coupling to the surrounding $^{13}$C nuclear spin bath gives rise to the self-discharging of the QB. The microwave (MW) field driving the electronic spin and the radio-frequency (RF) field initializing the native $^{14}$N nuclear spin can improve the quantum coherence of the QB to mitigate the self-discharging. The former one serves as the direct charging and the latter one is mediated by the hyperfine coupling.}
    \label{Fig1}
\end{figure}

In this Letter, we propose a QB scheme based on the nitrogen-vacancy (NV) center in diamond, which has been widely used in quantum information processing \cite{doi:10.1126/science.1139831, Balasubramanian2008, Taylor2008, science.1181193, Fuchs2011, PhysRevX.5.041001, Humphreys2018, PhysRevX.9.031045, PhysRevLett.126.197702,RevModPhys.92.015004,PhysRevLett.130.133603}. Being realized by the electronic spin of the NV center, the QB can be directly charged by an external field. The unique hyperfine coupling of the QB to the native $^{14}$N nucleus offers a useful coherent-control channel for the QB. Via investigating the impacts of decoherence caused by the surrounding $^{13}$C nuclear spin bath on the QB, we find that the coherent ergotropy decays more slowly than the incoherent ergotropy under the self-discharging effect. Therefore, the improvement the ratio of coherent ergotropy to total ergotropy during the charging process is advisable to enhance the inherent robustness of the QB to the self-discharging effect in the storage process. Furthermore, we find that a $100\%$ ratio of coherent ergotropy to total ergotropy can be obtained both in the transient state and the steady state of the QB, which can be optimized as long as the quantum coherence of the QB is maximized. These results can deepen the understanding of extractable work and enrich the theoretical references for the realization of QBs.

\section{QB scheme}
We propose to realize a QB based on the NV center in diamond, as shown in Fig. \ref{Fig1}. It has an electronic spin with $S=1$ and a native $^{14}$N nuclear spin with $I=1$. Its Hamiltonian reads \cite{Neumann2010,PhysRevLett.121.060401,Yang_2019}
\begin{equation}
\hat{H}_s=D\hat{S}_{z}^2+\gamma_e\mathbf{B}\cdot\mathbf{\hat{S}}-Q\hat{I}_{z}^2-\gamma_n\mathbf{B}\cdot\mathbf{\hat{I}}-\mathbf{\hat{S}}\cdot\mathbf{\bar{A}}\cdot\mathbf{\hat{I}}.\label{orghmt}
\end{equation}
Here, $\mathbf{\hat{S}}=(\hat{S}_{x},\hat{S}_{y},\hat{S}_{z})$ is the electronic spin operator with the zero-field splitting $D/2\pi=2.87$ GHz and the gyromagnetic ratio $\gamma_e/2\pi=2.8$ MHz/G. $\mathbf{\hat{I}}=(\hat{I}_{x},\hat{I}_{y}, \hat{I}_{z})$ is the $^{14}$N nuclear spin operator with the zero-field splitting $Q/2\pi=4.96$ MHz and the gyromagnetic ratio $\gamma_n/2\pi=3.07\times10^{-4}$ MHz/G. $\mathbf{\bar{A}}=\text{diag}(A_\perp,A_\perp,A_\parallel)$, with $A_\perp/2\pi=2.7$ MHz and $A_\parallel/2\pi=2.14$ MHz, is the hyperfine coupling strength between the electronic spin and the $^{14}$N nuclear spin. Assuming the magnetic field $\mathbf{B}=(0,0,B_z)$ and using the rotating-wave approximation, one can find an isolated subspace spanned by four bases $\vert m_S=0,-1\rangle\otimes\vert m_I=0,+1\rangle$. The electronic spin is taken as the QB, whose exhausted state $\vert g\rangle$ and fully charged state $\vert e\rangle$ are encoded in $\vert m_S=0\rangle$ and $\vert m_S=-1\rangle$, respectively. In this subspace, Eq. \eqref{orghmt} reduces to $\hat{H}_s\simeq \hat{H}_b+\hat{H}_\text{hf}$, with $\hat{H}_b=\omega_0\hat{\sigma}^\dagger\hat{\sigma}$ and $\hat{H}_\text{hf}=A_\parallel\hat{\sigma}^\dagger\hat{\sigma}\hat{h}^\dagger\hat{h}$, where $\omega_0=D-\gamma_e B_z$, $\hat{\sigma}=\vert g\rangle \langle e\vert$, and $\hat{h}=\left\vert\downarrow\right\rangle\left\langle\uparrow\right\vert$, with $\left\vert\uparrow\right\rangle=\vert m_I=+1\rangle$ and $\left\vert\downarrow\right\rangle=\vert m_I=0\rangle$. The charging of the QB is realized by applying an external microwave (MW) field to the NV center. The charging Hamiltonian is
\begin{equation}
    \hat{H}_c=\frac{\Omega}{2}(\hat{\sigma}e^{i\omega t}+\text{H.c.}),\label{HC}
\end{equation}
where $\Omega$ is the driving strength and $\omega$ is the driving frequency. Additionally, as the state of the native $^{14}$N nuclear spin can be initialized by exerting an external rf field on the NV center, the unique hyperfine coupling offers a useful channel to coherently control the QB, which has not been considered in the previous schemes.

Naturally, the NV center in diamond also has randomly distributed $^{13}$C nuclear spins with abundance $1.1\%$. They act as a bath to the electronic spin. The weak incoherent interaction between the electronic spin and this $^{13}$C spin bath causes the QB to lose its energy irreversibly, resulting in an uncontrollable self-discharging of our QB. Therefore, before connecting the QB to a consumption hub, its dynamics is governed by the Born-Markov master equation \cite{PhysRevLett.124.210502}
\begin{equation}
    \dot{\rho}(t)=-i[\hat{H}_{eff},\rho(t)]+\frac{\gamma}{2}[2\hat{\sigma}\rho(t)\hat{\sigma}^\dagger-\{\hat{\sigma}^\dagger\hat{\sigma},\rho(t)\}],\label{ME}
\end{equation}
with $\hat{H}_\text{eff}=\Delta\hat{\sigma}^\dagger\hat{\sigma}+(\Omega/2)\hat{\sigma}_x+A_\parallel\hat{\sigma}^\dagger\hat{\sigma}\hat{h}^\dagger\hat{h}$. $\rho(t)$ is the total density matrix of the QB and the native $^{14}$N nuclear spin. $\Delta=\omega_0-\omega$ is the frequency detuning. $\gamma$ is the decay rate of electronic spin, which is dependent on the concentration of $^{13}$C nuclear spins \cite{DOHERTY20131}. An ultralong spin coherence time can be obtained in isotopically engineered diamond \cite{Balasubramanian2009}. 

A basic quantity to characterize the performance of the QB is the stored energy $\mathcal{E}$. It is defined as $\mathcal{E}(t)=\text{Tr}[\hat{H}_b\rho_b(t)]$, with $\rho_b(t)$ being the reduced density matrix of the QB. However, constrained by the second law of thermodynamics, not all of the stored energy can be converted into work. A key quantity called the ergotropy quantifies the QB's maximum amount of extractable work by the unitary dynamics. It is defined as $\mathcal{W}(t)=\text{Tr}[\hat{H}_b\rho_b(t)]-\text{Tr}[\hat{H}_b\tilde\rho_b(t)]$, where $\tilde\rho_b(t)=\sum_{k}r_k(t)\vert\varepsilon_k\rangle\langle\varepsilon_k\vert$ is the passive counterpart of $\rho_b(t)$, with $r_k(t)$ being the eigenvalues of $\rho_b(t)$ ordered in a descending sort and $\vert\varepsilon_k\rangle$ being the eigenstates of $\hat{H}_b$ with the corresponding eigenvalues $\varepsilon_k$ ordered in an ascending sort \cite{A.E.Allahverdyan_2004}. Quantum coherence has been found to play a significant role in work extraction from quantum systems \cite{PhysRevE.103.042118,PhysRevLett.125.180603,PhysRevLett.131.060402}. It is quantified by
$\mathcal{C}(t)=\mathcal{S}[\varrho_b(t)]-\mathcal{S}[\rho_b(t)]$, where $\varrho_b(t)=\sum_{k}\langle\varepsilon_k\vert\rho_b(t)\vert\varepsilon_k\rangle\vert\varepsilon_k\rangle\langle\varepsilon_k\vert$ is the dephased counterpart of $\rho_b(t)$ and $\mathcal{S}[\rho]=-\text{Tr}[\rho\log\rho]$ is the Von Neumann entropy of the corresponding density matrix \cite{PhysRevLett.113.140401}. Quantum coherence permits to separate ergotropy into the coherent and incoherent parts \cite{PhysRevLett.125.180603,PhysRevLett.129.130602}. Characterizing the maximum work unitarily extractable from $\rho_b(t)$ without altering its quantum coherence, the incoherent ergotropy $\mathcal{W}^i(t)$ is defined as
\begin{equation}
    \mathcal{W}^i(t)=\text{Tr}[\hat{H}_b\rho_b(t)]-\text{Tr}[\hat{H}_b\tilde{\varrho}_b(t)],
\end{equation}
where $\tilde{\varrho}_b(t)=\sum_{k}\lambda_k(t)\vert\varepsilon_k\rangle\langle\varepsilon_k\vert$ is the passive counterpart of the dephased state $\varrho_b(t)$. Denoting the extractable work due to the
presence of the QB's quantum coherence, the coherent ergotropy $\mathcal{W}^c$ reads
\begin{equation}
    \mathcal{W}^c(t)=\text{Tr}[\hat{H}_b\tilde{\varrho}_b(t)]-\text{Tr}[\hat{H}_b\tilde\rho_b(t)].
\end{equation}
It is easy to verify that $\mathcal{W}^c=0$ when the QB does not have quantum coherence. This inspires us to increase the coherent ergotropy of the QB by optimizing its quantum coherence, which is realizable by the MW field and the hyperfine interaction.  

\section{QB performance} 
\begin{figure}[t]
    \centering
    \includegraphics[width=0.95\linewidth]{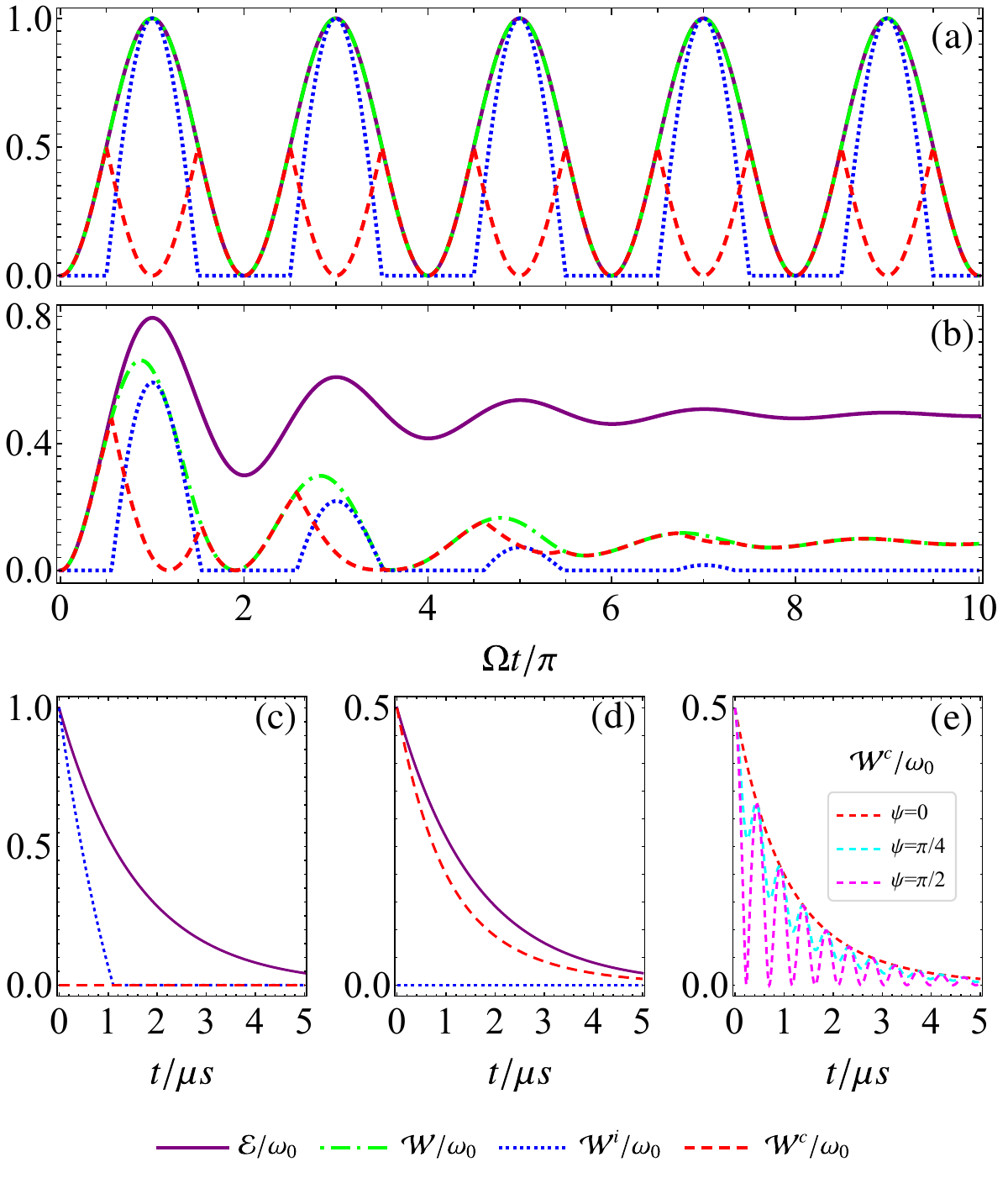}
    \caption{Evolution of the stored energy $\mathcal{E}$, the ergotropy $\mathcal{W}$, the incoherent ergotropy $\mathcal{W}^i$, and the coherent ergotropy $\mathcal{W}^c$ in the charging process with $\left\vert\Psi(0)\right\rangle=\left\vert g\right\rangle\otimes\left\vert\downarrow\right\rangle$, $\Omega/2\pi=0.5$ MHz, $\Delta=0$ when (a) $\gamma=0$, and (b) $\gamma/2\pi=0.1$ MHz. Evolution of $\mathcal{E}$, $\mathcal{W}^i$, and $\mathcal{W}^c$ in the storage process with $\gamma/2\pi=0.1$ MHz when the charging process ends at $\Omega t=\pi$ in (c) and $\pi/2$ in (d). (e) Evolution of $\mathcal{W}^c$ in the storage process with $\gamma/2\pi=0.1$ MHz when the charging process ends at $\Omega t=\pi/2$ when the initial state $\psi=0$ (red dashed line), $\pi/4$ (cyan dashed line), and $\pi/2$ (magenta dashed line).}
    \label{Fig2}
\end{figure}

In our QB scheme, there are one decoherence channel and two coherent-control channels. The decoherence channel is induced by the incoherent coupling to the spin bath of $^{13}$C nuclei, which causes the self-discharging of the QB. The coherent-control channels are provided by the external MW field applied to the electronic spin and the hyperfine interaction with the native $^{14}$N nuclear spin, which can enhance the quantum coherence of the QB. In order to figure out the interplay between the decoherence channel and the coherent-control channels, the dynamical behaviors of incoherent and coherent ergotropies during the charging process and the storage process are shown in Fig. \ref{Fig2}. During the charging process, the initial state is $|\Psi(0)\rangle=|g\rangle\otimes\left\vert\downarrow\right\rangle$, under which the QB and the native $^{14}$N nuclear spin are decoupled and the effective Hamiltonian can be reduced to $\hat{H}_\text{eff}=\Delta\hat{\sigma}^\dagger\hat{\sigma}+(\Omega/2)\hat{\sigma}_x$. Figure \ref{Fig2}(a) shows the charging dynamics in the absence of self-discharging when $\gamma=0$. The time-dependent state of the QB and the native $^{14}$N nuclear spin is $\vert\Psi(t)\rangle=[\sin(\Omega t/2)\vert e\rangle+\cos(\Omega t/2)\vert g\rangle]\otimes\left\vert\downarrow\right\rangle$. In this case, $\mathcal{E}(t)=\mathcal{W}(t)=\omega_0\sin^2(\Omega t/2)$, both of which oscillate with a period $2\pi/\Omega$ over time. It indicates that the energy of the QB obtained through this charging channel is totally convertible into work by unitary operations. $\mathcal{W}^c(t)$ and $\mathcal{W}^i(t)$, one of which increases and the other decreases, evolve inversely with time. Consistent with the result illustrated in Ref. \cite{PhysRevLett.129.130602}, $\mathcal{W}^c(t)$ reaches its maximum earlier than $\mathcal{W}^i(t)$. When the charging process ends at $\Omega t=\pi$, the QB is fully charged but its quantum coherence is zero, thus the coherent ergotropy is zero. When the charging process ends at $\Omega t=\pi/2$, the QB is half charged but its quantum coherence is $1$, thus the coherent ergotropy is maximal. Figure \ref{Fig2}(b) shows the charging dynamics in the presence of the self-discharging. The charging performance is deteriorated by the self-discharging. In the long-time limit, the QB evolves to the steady state \begin{equation}
    \rho_b(\infty)=\frac{1}{\eta}\left(\begin{array}{cc}
        \Omega^2 & -2\Delta\Omega-i\gamma\Omega \\
        -2\Delta\Omega+i\gamma\Omega & \eta-\Omega^2
    \end{array}\right).
\end{equation}
with $\eta=4\Delta^2+2\Omega^2+\gamma^2$. 
In $\rho_b(\infty)$, the stored energy of the QB is $\mathcal{E}(\infty)=\omega_0\Omega^2/\eta$ and the ratio of $\mathcal{W}^c(\infty)$ to $\mathcal{W}(\infty)=\omega_0\sqrt{4\Delta^2+\gamma^2}(\sqrt{\eta+2\Omega^2}-\sqrt{4\Delta^2+\gamma^2})/\eta$ is $100\%$. It indicates the robustness of the coherent ergotropy and the fragility of the incoherent ergotropy to the self-discharging.

Next, we evaluate the effect of the hyperfine interaction on the energy-storage performance of the QB. The state of $^{14}$N nuclear spin is assumed to be initialized into $|\psi\rangle=\sin(\psi/2)\left\vert\uparrow\right\rangle+\cos(\psi/2)\left\vert\downarrow\right\rangle$ by the rf field. During the storage process, the MW field is turned off, thus $\Omega=0$ and $\Delta=0$. The effective Hamiltonian can be simplified into $\hat{H}_\text{eff}=A_\parallel\hat{\sigma}^\dagger\hat{\sigma}\hat{h}^\dagger\hat{h}$. Figures \ref{Fig2}(c) and \ref{Fig2}(d) show that $\mathcal{E}(t)$ and $\mathcal{W}^{i,c}(t)$ irreversibly decrease to zero due to the uncontrollable self-discharging no matter what the value of $\psi$ is. If the QB is fully charged at the initial time of the storage process, then the coherent ergotropy remains zero and thus we do not have space to optimize both the coherent and incoherent ergotropies via the hyperfine interaction; see Fig. \ref{Fig2}(c). On the contrary, if the QB is half charged at the initial time of the storage process, $\mathcal{W}^{i}$ remains zero while $\mathcal{W}^c$ can be controlled by modulating the initial state of the $^{14}$N nuclear spin via $\psi$; see Figs. \ref{Fig2}(d) and \ref{Fig2}(e). This feature matches the non-Markovian behavior presented in Ref. \cite{PhysRevLett.124.210502} and provides us a useful coherent-control channel for the QB. Compared with the behavior of  suddenly damping to zero of $\mathcal{W}^i(t)$ in Fig. \ref{Fig2}(c), $\mathcal{W}^c(t)$ in Fig. \ref{Fig2}(d) decays more slowly. Therefore, the improvement of the ratio of $\mathcal{W}^c$ to $\mathcal{W}$ during the charging process is helpful to enhancing the robustness of the QB to the self-discharging during the storage process.

\begin{figure}[t]
    \centering
    \includegraphics[width=\columnwidth]{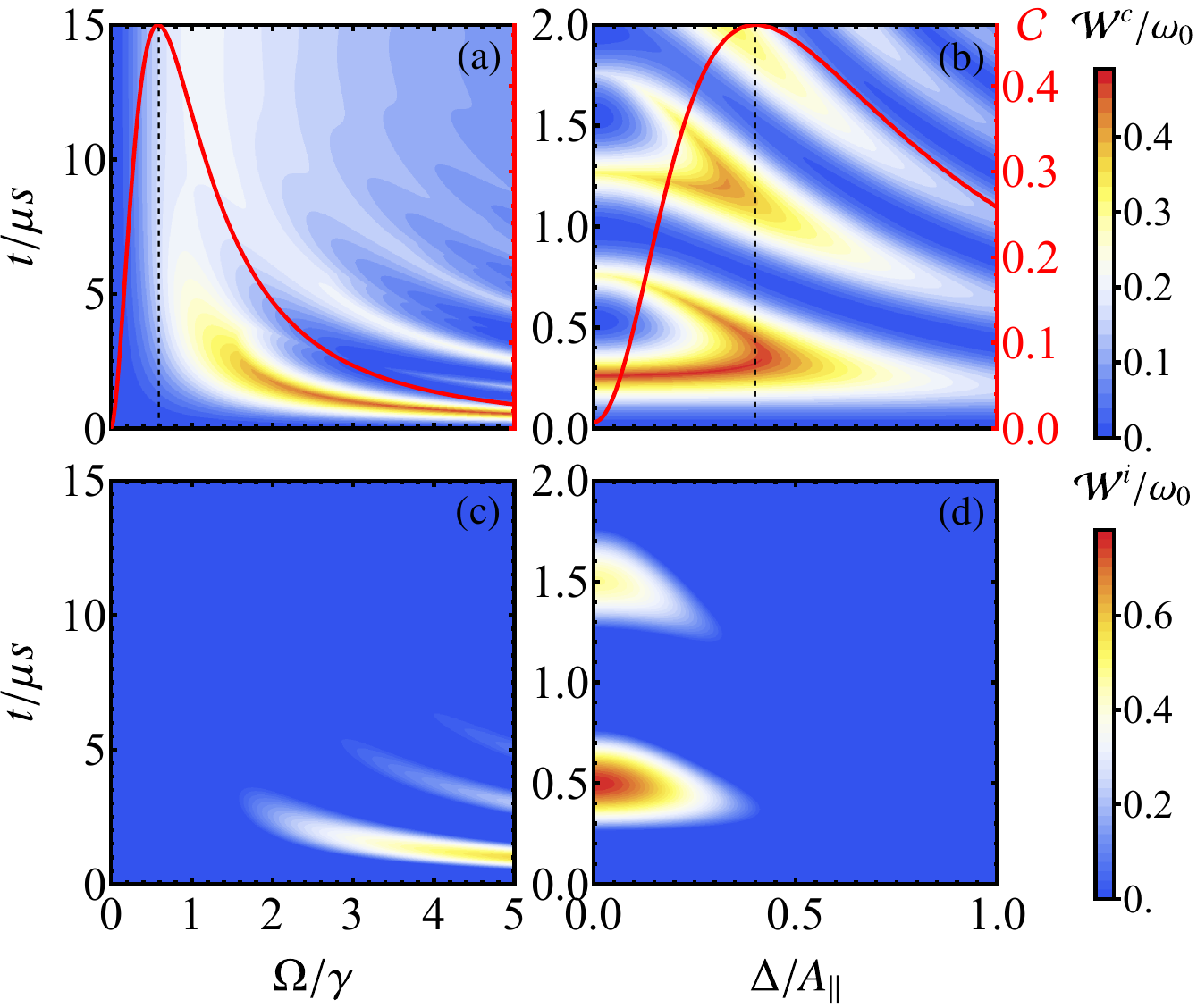}
    \caption{Dynamic behaviors of (a), (b) the coherent ergotropy $\mathcal{W}^c$ and (c), (d) the incoherent ergotropy $\mathcal{W}^i$ changing with (a), (c) the driving strength $\Omega$ when $\Delta=0$ and (b), (d) the frequency detuning $\Delta$ when $\Omega/2\pi=1$ MHz. The corresponding stable behavior of the quantum coherence $\mathcal{C}$ is described by the red solid lines overlaid on the density plots in (a), (b). Other parameters used for numerical simulation are $\gamma/2\pi=0.1$ MHz and $\vert\Psi(0)\rangle=\vert g\rangle\otimes\left\vert\downarrow\right\rangle$.}
    \label{Fig3}
\end{figure}

\begin{figure}[b]
    \centering
    \includegraphics[width=\columnwidth]{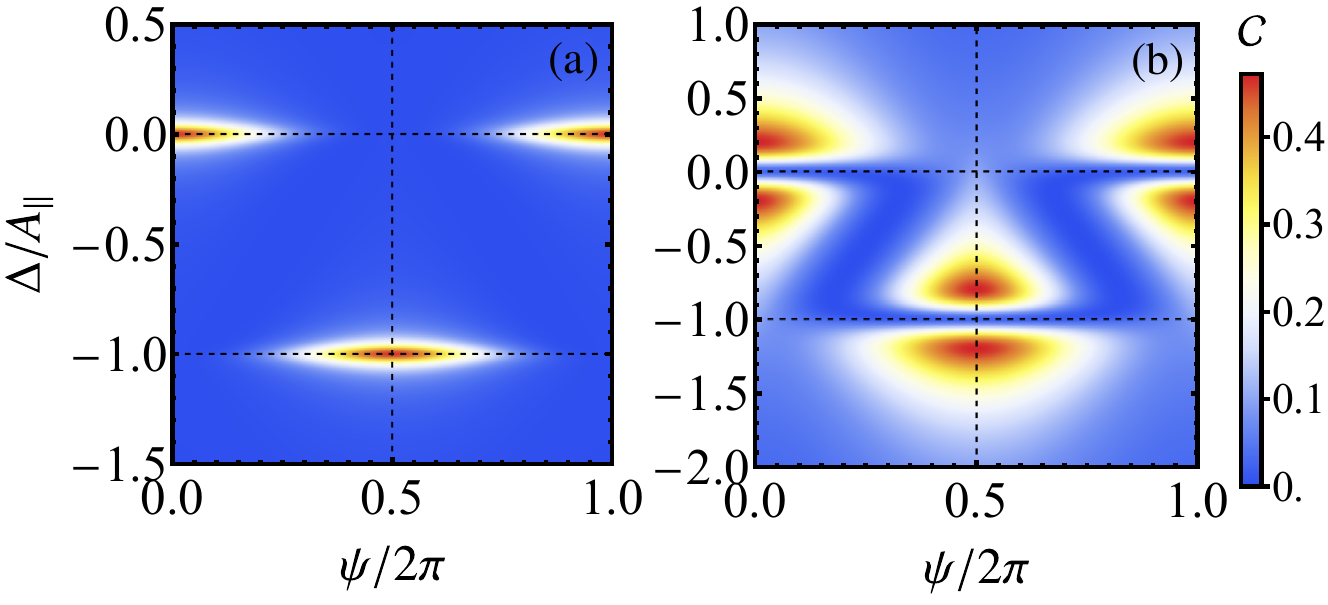}
    \caption{Stable behavior of the quantum coherence $\mathcal{C}$ changing with the initial phase $\psi$ and the frequency detuning $\Delta$ when (a) $\Omega/2\pi=0.05$ MHz and (b) $\Omega/2\pi=0.5$ MHz. Other parameters used for numerical simulation are $\gamma/2\pi=0.1$ MHz and $\vert\Psi(0)\rangle=\vert g\rangle\otimes\vert\psi\rangle$.}
    \label{Fig4}
\end{figure}

The coherent-control effect of the MW field on the coherent ergotropy is studied in Fig. \ref{Fig3}. Figures \ref{Fig3}(a) and \ref{Fig3}(c) show the evolution of $\mathcal{W}^c(t)$ and $\mathcal{W}^i(t)$ during the charging process in different driving strength $\Omega$. When $\Omega/\gamma<0.6$, $\mathcal{W}^c(t)$ monotonically increases with time and saturates to stable values, but $\mathcal{W}^i(t)=0$, accompanied by an increasing of $\mathcal{C}$ with $\Omega$. When $0.6<\Omega/\gamma<1.6$, $\mathcal{W}^c(t)$ begins to oscillate with time and $\mathcal{W}^i(t)=0$, but $\mathcal{C}$ begins to decrease with $\Omega$. When $\Omega/\gamma>1.6$, $\mathcal{W}^c(t)$ oscillates quickly with time and $\mathcal{W}^i(t)\ne0$, accompanied by a further decreasing of $\mathcal{C}$ with $\Omega$. This phenomenon indicates that a small driving strength is favorable to achieve a stable $\mathcal{W}^c$ and increase the ratio of $\mathcal{W}^c$ to $\mathcal{W}$. When $\Omega/\gamma=0.6$, the quantum coherence is maximal and the achieved stable $\mathcal{W}^c$ is optimal. This optimization is manifested in that the time required to reach the stable $\mathcal{W}^c$ is shorter and its stabilized value is larger than those given by $\Omega/\gamma<0.6$. Although a larger $\mathcal{W}^c$ is achievable in the larger values of $\Omega$, the rapid oscillation makes it hard to precisely control its time instant and the nonzero $\mathcal{W}^i$ reduces the ratio of $\mathcal{W}^c$ to $\mathcal{W}$. Figures \ref{Fig3}(b) and \ref{Fig3}(d) show the evolution of $\mathcal{W}^c(t)$ and $\mathcal{W}^i(t)$ during the charging process in different detuning $\Delta$. When $\Delta/A_\parallel<0.4$, the oscillation peak of $\mathcal{W}^c$ is narrow over time and $\mathcal{W}^i\ne0$, accompanied by an increasing of $\mathcal{C}$ with $\Delta$. When $\Delta/A_\parallel>0.4$, the oscillation peak of $\mathcal{W}^c$ becomes wide, $\mathcal{W}^i$ becomes zero, and $\mathcal{C}$ begins to decrease with $\Delta$. When $\Delta/A_\parallel=0.4$, the quantum coherence is maximal and the maximal coherent ergotropy is optimal. This optimization is manifested in that both the peak width and the peak value featured by the maximal coherent ergotropy are the largest and the incoherent ergotropy is zero. The widening of the peak of the maximal coherent ergotropy reduces the difficulty in addressing the time instant at which the coherent ergotropy is optimal. According to the above analysis, we can shorten the time required to reach the stable coherent ergotropy and widen its oscillation peak via maximizing the quantum coherence by changing the driving strength and the frequency detuning of the MW field.

The coherent-control effect of the hyperfine interaction on the quantum coherence is studied in Fig. \ref{Fig4}. The native nuclear spin is initialized in a superposition state $|\psi\rangle$ by the rf field.  $\mathcal{C}$ can be maximized by choosing proper values of $\psi$ and $\Delta$ for different $\Omega$. When $\Omega/2\pi=0.05$ MHz, the maximal quantum coherence can be achieved by setting $\Delta=0$ for the initial state $\vert\psi\rangle=\left\vert\downarrow\right\rangle$ and $\Delta=-A_\parallel$ for the initial state $\vert\psi\rangle=\left\vert\uparrow\right\rangle$. It indicates that charging by resonant driving field is helpful to improve the QB's quantum coherence when the native nuclear spin is decoupled and the energy shift induced by the hyperfine coupling can be compensated by the off-resonant driving. When $\Omega/2\pi=0.5$ MHz, each optimal point that appears in the situation of $\Omega/2\pi=0.05$ MHz is split by a slight frequency detuning. It reflects that detuning driving field is more advantageous for improving the QB's quantum coherence when the driving strength is large. Thus, the advantage of our QB is that its unique hyperfine interaction with the $^{14}$N nucleus permits us to coherently enhance the quantum coherence and thus optimize the ratio of coherent ergotropy to total ergotropy, which gives us with the ability to mitigate the self-discharging on one hand and to maximize the extractable work on the other.

\section{Discussion and conclusion}
Recently, it has been observed that quantum coherence promotes the coherent ergotropy in the electronic spin of the NV center, where the coherent and incoherent ergotropies are successfully measured \cite{PhysRevLett.133.180401}. This advance provides technical support for the realization of our QB scheme. In a magnetic field with $B_z=482$ G, the QB energy $\omega_0$ is about $6.28\ \mu$eV. The time required to fully charge it depends on the driving strength $\Omega/2\pi$, which may range from $0.1$ to $24$ MHz. The time duration that the QB can store energy is determined by the decay rate $\gamma/2\pi$, whose value ranges from $0.01$ to $0.34$ MHz \cite{PhysRevLett.124.210502,PhysRevLett.133.180401}. Under $\Omega/2\pi=1$ MHz and $\gamma/2\pi=0.1$ MHz, the incoherent ergotropy reaches its maximum $6.28\ \mu$eV at $t=0.5\ \mu$s in the charging process, where both the quantum coherence and the coherent-to-total ergotropy ratio are zero. Then, the ergotropy remains only for $1\ \mu$s. In contrast, the coherent ergotropy reaches its maximum $3.14\ \mu$eV after spending $t=0.25\ \mu$s in the charging process, where the quantum coherence becomes maximal and the coherent-to-total ergotropy ratio reaches 100\%. In this case, the ergotropy can remain for about $8\ \mu$s in the storage process. The storage time of the ergotropy is enhanced 8 times by optimizing the quantum coherence and the coherent-to-total ergotropy ratio. The hyperfine interaction supplies a useful way to control the quantum coherence, which can be realized by tuning the initial state $|\psi\rangle$ of the $^{14}$N via manipulating the duration and strength of the applied rf field. 

In conclusion, we have proposed a QB scheme based on the NV center in diamond. The electronic spin state of the NV center serves as the QB, which can be charged by the MW field. Exploiting the finding that the coherent ergotropy is more robust to the self-discharging than the incoherent ergotropy, we have identified a mechanism to mitigate the self-discharging during the storage process by maximizing the ratio of coherent ergotropy to total ergotropy. The unique hyperfine interaction between the electronic spin and the $^{14}$N nuclear spin provides a valuable coherent-control channel for optimizing the coherent ergotropy. Our result enriches the realization of the QB and lays the foundation for overcoming the self-discharging issue caused by various types of decoherence in practical applications. By unifying theoretical insight with an experimentally feasible system, we provide a definitive step toward realizing quantum energy devices. It is expected that, in close alignment with recent experimental progress \cite{PhysRevLett.124.210502,PhysRevLett.133.180401}, our work would hopefully prompt the development of NV-center quantum thermodynamics.

This work is supported by the National Natural Science Foundation of China (Grants No. 12105089,  No. 12074107, No. 12275109, No. 12274422, and No. 12247101), the Innovation Program for Quantum Science and Technology (Grant No. 2023ZD0300400), the innovation group project of Hubei Province (Grant No. 2022CFA012), the program of outstanding young and middle-aged scientific and technological innovation team of colleges and universities in Hubei Province (Grant No. T2020001), and the Hubei Province Science Fund for Distinguished Young Scholars (Grant No. 2020CFA078).

\bibliography{reference}

\begin{thebibliography}{85}%
\makeatletter
\providecommand \@ifxundefined [1]{%
 \@ifx{#1\undefined}
}%
\providecommand \@ifnum [1]{%
 \ifnum #1\expandafter \@firstoftwo
 \else \expandafter \@secondoftwo
 \fi
}%
\providecommand \@ifx [1]{%
 \ifx #1\expandafter \@firstoftwo
 \else \expandafter \@secondoftwo
 \fi
}%
\providecommand \natexlab [1]{#1}%
\providecommand \enquote  [1]{``#1''}%
\providecommand \bibnamefont  [1]{#1}%
\providecommand \bibfnamefont [1]{#1}%
\providecommand \citenamefont [1]{#1}%
\providecommand \href@noop [0]{\@secondoftwo}%
\providecommand \href [0]{\begingroup \@sanitize@url \@href}%
\providecommand \@href[1]{\@@startlink{#1}\@@href}%
\providecommand \@@href[1]{\endgroup#1\@@endlink}%
\providecommand \@sanitize@url [0]{\catcode `\\12\catcode `\$12\catcode
  `\&12\catcode `\#12\catcode `\^12\catcode `\_12\catcode `\%12\relax}%
\providecommand \@@startlink[1]{}%
\providecommand \@@endlink[0]{}%
\providecommand \url  [0]{\begingroup\@sanitize@url \@url }%
\providecommand \@url [1]{\endgroup\@href {#1}{\urlprefix }}%
\providecommand \urlprefix  [0]{URL }%
\providecommand \Eprint [0]{\href }%
\providecommand \doibase [0]{https://doi.org/}%
\providecommand \selectlanguage [0]{\@gobble}%
\providecommand \bibinfo  [0]{\@secondoftwo}%
\providecommand \bibfield  [0]{\@secondoftwo}%
\providecommand \translation [1]{[#1]}%
\providecommand \BibitemOpen [0]{}%
\providecommand \bibitemStop [0]{}%
\providecommand \bibitemNoStop [0]{.\EOS\space}%
\providecommand \EOS [0]{\spacefactor3000\relax}%
\providecommand \BibitemShut  [1]{\csname bibitem#1\endcsname}%
\let\auto@bib@innerbib\@empty
\bibitem [{\citenamefont {\ifmmode \acute{C}\else
  \'{C}\fi{}wikli\ifmmode~\acute{n}\else \'{n}\fi{}ski}\ \emph
  {et~al.}(2015)\citenamefont {\ifmmode \acute{C}\else
  \'{C}\fi{}wikli\ifmmode~\acute{n}\else \'{n}\fi{}ski}, \citenamefont
  {Studzi\ifmmode~\acute{n}\else \'{n}\fi{}ski}, \citenamefont {Horodecki},\
  and\ \citenamefont {Oppenheim}}]{PhysRevLett.115.210403}%
  \BibitemOpen
  \bibfield  {author} {\bibinfo {author} {\bibfnamefont {P.}~\bibnamefont
  {\ifmmode \acute{C}\else \'{C}\fi{}wikli\ifmmode~\acute{n}\else
  \'{n}\fi{}ski}}, \bibinfo {author} {\bibfnamefont {M.}~\bibnamefont
  {Studzi\ifmmode~\acute{n}\else \'{n}\fi{}ski}}, \bibinfo {author}
  {\bibfnamefont {M.}~\bibnamefont {Horodecki}},\ and\ \bibinfo {author}
  {\bibfnamefont {J.}~\bibnamefont {Oppenheim}},\ }\bibfield  {title} {\bibinfo
  {title} {{Limitations on the Evolution of Quantum Coherences: Towards Fully
  Quantum Second Laws of Thermodynamics}},\ }\href
  {https://doi.org/10.1103/PhysRevLett.115.210403} {\bibfield  {journal}
  {\bibinfo  {journal} {Phys. Rev. Lett.}\ }\textbf {\bibinfo {volume} {115}},\
  \bibinfo {pages} {210403} (\bibinfo {year} {2015})}\BibitemShut {NoStop}%
\bibitem [{\citenamefont {Bera}\ \emph {et~al.}(2017)\citenamefont {Bera},
  \citenamefont {Riera}, \citenamefont {Lewenstein},\ and\ \citenamefont
  {Winter}}]{Bera2017}%
  \BibitemOpen
  \bibfield  {author} {\bibinfo {author} {\bibfnamefont {M.~N.}\ \bibnamefont
  {Bera}}, \bibinfo {author} {\bibfnamefont {A.}~\bibnamefont {Riera}},
  \bibinfo {author} {\bibfnamefont {M.}~\bibnamefont {Lewenstein}},\ and\
  \bibinfo {author} {\bibfnamefont {A.}~\bibnamefont {Winter}},\ }\bibfield
  {title} {\bibinfo {title} {Generalized laws of thermodynamics in the presence
  of correlations},\ }\href {https://doi.org/10.1038/s41467-017-02370-x}
  {\bibfield  {journal} {\bibinfo  {journal} {Nature Communications}\ }\textbf
  {\bibinfo {volume} {8}},\ \bibinfo {pages} {2180} (\bibinfo {year}
  {2017})}\BibitemShut {NoStop}%
\bibitem [{\citenamefont {Alicki}\ and\ \citenamefont
  {Kosloff}(2018)}]{Alicki2018}%
  \BibitemOpen
  \bibfield  {author} {\bibinfo {author} {\bibfnamefont {R.}~\bibnamefont
  {Alicki}}\ and\ \bibinfo {author} {\bibfnamefont {R.}~\bibnamefont
  {Kosloff}},\ }\bibinfo {title} {Introduction to quantum thermodynamics:
  History and prospects},\ in\ \href
  {https://doi.org/10.1007/978-3-319-99046-0_1} {\emph {\bibinfo {booktitle}
  {Thermodynamics in the Quantum Regime: Fundamental Aspects and New
  Directions}}}\ (\bibinfo  {publisher} {Springer International Publishing},\
  \bibinfo {address} {Cham},\ \bibinfo {year} {2018})\ pp.\ \bibinfo {pages}
  {1--33}\BibitemShut {NoStop}%
\bibitem [{\citenamefont {Ro\ss{}nagel}\ \emph {et~al.}(2014)\citenamefont
  {Ro\ss{}nagel}, \citenamefont {Abah}, \citenamefont {Schmidt-Kaler},
  \citenamefont {Singer},\ and\ \citenamefont {Lutz}}]{PhysRevLett.112.030602}%
  \BibitemOpen
  \bibfield  {author} {\bibinfo {author} {\bibfnamefont {J.}~\bibnamefont
  {Ro\ss{}nagel}}, \bibinfo {author} {\bibfnamefont {O.}~\bibnamefont {Abah}},
  \bibinfo {author} {\bibfnamefont {F.}~\bibnamefont {Schmidt-Kaler}}, \bibinfo
  {author} {\bibfnamefont {K.}~\bibnamefont {Singer}},\ and\ \bibinfo {author}
  {\bibfnamefont {E.}~\bibnamefont {Lutz}},\ }\bibfield  {title} {\bibinfo
  {title} {{Nanoscale Heat Engine Beyond the Carnot Limit}},\ }\href
  {https://doi.org/10.1103/PhysRevLett.112.030602} {\bibfield  {journal}
  {\bibinfo  {journal} {Phys. Rev. Lett.}\ }\textbf {\bibinfo {volume} {112}},\
  \bibinfo {pages} {030602} (\bibinfo {year} {2014})}\BibitemShut {NoStop}%
\bibitem [{\citenamefont {Klatzow}\ \emph {et~al.}(2019)\citenamefont
  {Klatzow}, \citenamefont {Becker}, \citenamefont {Ledingham}, \citenamefont
  {Weinzetl}, \citenamefont {Kaczmarek}, \citenamefont {Saunders},
  \citenamefont {Nunn}, \citenamefont {Walmsley}, \citenamefont {Uzdin},\ and\
  \citenamefont {Poem}}]{PhysRevLett.122.110601}%
  \BibitemOpen
  \bibfield  {author} {\bibinfo {author} {\bibfnamefont {J.}~\bibnamefont
  {Klatzow}}, \bibinfo {author} {\bibfnamefont {J.~N.}\ \bibnamefont {Becker}},
  \bibinfo {author} {\bibfnamefont {P.~M.}\ \bibnamefont {Ledingham}}, \bibinfo
  {author} {\bibfnamefont {C.}~\bibnamefont {Weinzetl}}, \bibinfo {author}
  {\bibfnamefont {K.~T.}\ \bibnamefont {Kaczmarek}}, \bibinfo {author}
  {\bibfnamefont {D.~J.}\ \bibnamefont {Saunders}}, \bibinfo {author}
  {\bibfnamefont {J.}~\bibnamefont {Nunn}}, \bibinfo {author} {\bibfnamefont
  {I.~A.}\ \bibnamefont {Walmsley}}, \bibinfo {author} {\bibfnamefont
  {R.}~\bibnamefont {Uzdin}},\ and\ \bibinfo {author} {\bibfnamefont
  {E.}~\bibnamefont {Poem}},\ }\bibfield  {title} {\bibinfo {title}
  {{Experimental Demonstration of Quantum Effects in the Operation of
  Microscopic Heat Engines}},\ }\href
  {https://doi.org/10.1103/PhysRevLett.122.110601} {\bibfield  {journal}
  {\bibinfo  {journal} {Phys. Rev. Lett.}\ }\textbf {\bibinfo {volume} {122}},\
  \bibinfo {pages} {110601} (\bibinfo {year} {2019})}\BibitemShut {NoStop}%
\bibitem [{\citenamefont {Micadei}\ \emph {et~al.}(2019)\citenamefont
  {Micadei}, \citenamefont {Peterson}, \citenamefont {Souza}, \citenamefont
  {Sarthour}, \citenamefont {Oliveira}, \citenamefont {Landi}, \citenamefont
  {Batalh{\~a}o}, \citenamefont {Serra},\ and\ \citenamefont
  {Lutz}}]{Micadei2019}%
  \BibitemOpen
  \bibfield  {author} {\bibinfo {author} {\bibfnamefont {K.}~\bibnamefont
  {Micadei}}, \bibinfo {author} {\bibfnamefont {J.~P.~S.}\ \bibnamefont
  {Peterson}}, \bibinfo {author} {\bibfnamefont {A.~M.}\ \bibnamefont {Souza}},
  \bibinfo {author} {\bibfnamefont {R.~S.}\ \bibnamefont {Sarthour}}, \bibinfo
  {author} {\bibfnamefont {I.~S.}\ \bibnamefont {Oliveira}}, \bibinfo {author}
  {\bibfnamefont {G.~T.}\ \bibnamefont {Landi}}, \bibinfo {author}
  {\bibfnamefont {T.~B.}\ \bibnamefont {Batalh{\~a}o}}, \bibinfo {author}
  {\bibfnamefont {R.~M.}\ \bibnamefont {Serra}},\ and\ \bibinfo {author}
  {\bibfnamefont {E.}~\bibnamefont {Lutz}},\ }\bibfield  {title} {\bibinfo
  {title} {Reversing the direction of heat flow using quantum correlations},\
  }\href {https://doi.org/10.1038/s41467-019-10333-7} {\bibfield  {journal}
  {\bibinfo  {journal} {Nature Communications}\ }\textbf {\bibinfo {volume}
  {10}},\ \bibinfo {pages} {2456} (\bibinfo {year} {2019})}\BibitemShut
  {NoStop}%
\bibitem [{\citenamefont {Mayer}\ \emph {et~al.}(2023)\citenamefont {Mayer},
  \citenamefont {Lutz},\ and\ \citenamefont {Widera}}]{Mayer2023}%
  \BibitemOpen
  \bibfield  {author} {\bibinfo {author} {\bibfnamefont {D.}~\bibnamefont
  {Mayer}}, \bibinfo {author} {\bibfnamefont {E.}~\bibnamefont {Lutz}},\ and\
  \bibinfo {author} {\bibfnamefont {A.}~\bibnamefont {Widera}},\ }\bibfield
  {title} {\bibinfo {title} {{Generalized Clausius inequalities in a
  nonequilibrium cold-atom system}},\ }\href
  {https://doi.org/10.1038/s42005-023-01175-3} {\bibfield  {journal} {\bibinfo
  {journal} {Communications Physics}\ }\textbf {\bibinfo {volume} {6}},\
  \bibinfo {pages} {61} (\bibinfo {year} {2023})}\BibitemShut {NoStop}%
\bibitem [{\citenamefont {Alicki}(1979)}]{Alicki_1979}%
  \BibitemOpen
  \bibfield  {author} {\bibinfo {author} {\bibfnamefont {R.}~\bibnamefont
  {Alicki}},\ }\bibfield  {title} {\bibinfo {title} {The quantum open system as
  a model of the heat engine},\ }\href
  {https://doi.org/10.1088/0305-4470/12/5/007} {\bibfield  {journal} {\bibinfo
  {journal} {Journal of Physics A: Mathematical and General}\ }\textbf
  {\bibinfo {volume} {12}},\ \bibinfo {pages} {L103} (\bibinfo {year}
  {1979})}\BibitemShut {NoStop}%
\bibitem [{\citenamefont {Linden}\ \emph {et~al.}(2010)\citenamefont {Linden},
  \citenamefont {Popescu},\ and\ \citenamefont
  {Skrzypczyk}}]{PhysRevLett.105.130401}%
  \BibitemOpen
  \bibfield  {author} {\bibinfo {author} {\bibfnamefont {N.}~\bibnamefont
  {Linden}}, \bibinfo {author} {\bibfnamefont {S.}~\bibnamefont {Popescu}},\
  and\ \bibinfo {author} {\bibfnamefont {P.}~\bibnamefont {Skrzypczyk}},\
  }\bibfield  {title} {\bibinfo {title} {{How Small Can Thermal Machines Be?
  The Smallest Possible Refrigerator}},\ }\href
  {https://doi.org/10.1103/PhysRevLett.105.130401} {\bibfield  {journal}
  {\bibinfo  {journal} {Phys. Rev. Lett.}\ }\textbf {\bibinfo {volume} {105}},\
  \bibinfo {pages} {130401} (\bibinfo {year} {2010})}\BibitemShut {NoStop}%
\bibitem [{\citenamefont {Roßnagel}\ \emph {et~al.}(2016)\citenamefont
  {Roßnagel}, \citenamefont {Dawkins}, \citenamefont {Tolazzi}, \citenamefont
  {Abah}, \citenamefont {Lutz}, \citenamefont {Schmidt-Kaler},\ and\
  \citenamefont {Singer}}]{doi:10.1126/science.aad6320}%
  \BibitemOpen
  \bibfield  {author} {\bibinfo {author} {\bibfnamefont {J.}~\bibnamefont
  {Roßnagel}}, \bibinfo {author} {\bibfnamefont {S.~T.}\ \bibnamefont
  {Dawkins}}, \bibinfo {author} {\bibfnamefont {K.~N.}\ \bibnamefont
  {Tolazzi}}, \bibinfo {author} {\bibfnamefont {O.}~\bibnamefont {Abah}},
  \bibinfo {author} {\bibfnamefont {E.}~\bibnamefont {Lutz}}, \bibinfo {author}
  {\bibfnamefont {F.}~\bibnamefont {Schmidt-Kaler}},\ and\ \bibinfo {author}
  {\bibfnamefont {K.}~\bibnamefont {Singer}},\ }\bibfield  {title} {\bibinfo
  {title} {A single-atom heat engine},\ }\href
  {https://doi.org/10.1126/science.aad6320} {\bibfield  {journal} {\bibinfo
  {journal} {Science}\ }\textbf {\bibinfo {volume} {352}},\ \bibinfo {pages}
  {325} (\bibinfo {year} {2016})}\BibitemShut {NoStop}%
\bibitem [{\citenamefont {Maslennikov}\ \emph {et~al.}(2019)\citenamefont
  {Maslennikov}, \citenamefont {Ding}, \citenamefont {Habl{\"u}tzel},
  \citenamefont {Gan}, \citenamefont {Roulet}, \citenamefont {Nimmrichter},
  \citenamefont {Dai}, \citenamefont {Scarani},\ and\ \citenamefont
  {Matsukevich}}]{Maslennikov2019}%
  \BibitemOpen
  \bibfield  {author} {\bibinfo {author} {\bibfnamefont {G.}~\bibnamefont
  {Maslennikov}}, \bibinfo {author} {\bibfnamefont {S.}~\bibnamefont {Ding}},
  \bibinfo {author} {\bibfnamefont {R.}~\bibnamefont {Habl{\"u}tzel}}, \bibinfo
  {author} {\bibfnamefont {J.}~\bibnamefont {Gan}}, \bibinfo {author}
  {\bibfnamefont {A.}~\bibnamefont {Roulet}}, \bibinfo {author} {\bibfnamefont
  {S.}~\bibnamefont {Nimmrichter}}, \bibinfo {author} {\bibfnamefont
  {J.}~\bibnamefont {Dai}}, \bibinfo {author} {\bibfnamefont {V.}~\bibnamefont
  {Scarani}},\ and\ \bibinfo {author} {\bibfnamefont {D.}~\bibnamefont
  {Matsukevich}},\ }\bibfield  {title} {\bibinfo {title} {Quantum absorption
  refrigerator with trapped ions},\ }\href
  {https://doi.org/10.1038/s41467-018-08090-0} {\bibfield  {journal} {\bibinfo
  {journal} {Nature Communications}\ }\textbf {\bibinfo {volume} {10}},\
  \bibinfo {pages} {202} (\bibinfo {year} {2019})}\BibitemShut {NoStop}%
\bibitem [{\citenamefont {Parrondo}\ \emph {et~al.}(2015)\citenamefont
  {Parrondo}, \citenamefont {Horowitz},\ and\ \citenamefont
  {Sagawa}}]{Parrondo2015}%
  \BibitemOpen
  \bibfield  {author} {\bibinfo {author} {\bibfnamefont {J.~M.~R.}\
  \bibnamefont {Parrondo}}, \bibinfo {author} {\bibfnamefont {J.~M.}\
  \bibnamefont {Horowitz}},\ and\ \bibinfo {author} {\bibfnamefont
  {T.}~\bibnamefont {Sagawa}},\ }\bibfield  {title} {\bibinfo {title}
  {Thermodynamics of information},\ }\href {https://doi.org/10.1038/nphys3230}
  {\bibfield  {journal} {\bibinfo  {journal} {Nature Physics}\ }\textbf
  {\bibinfo {volume} {11}},\ \bibinfo {pages} {131} (\bibinfo {year}
  {2015})}\BibitemShut {NoStop}%
\bibitem [{\citenamefont {Millen}\ and\ \citenamefont
  {Xuereb}(2016)}]{Millen_2016}%
  \BibitemOpen
  \bibfield  {author} {\bibinfo {author} {\bibfnamefont {J.}~\bibnamefont
  {Millen}}\ and\ \bibinfo {author} {\bibfnamefont {A.}~\bibnamefont
  {Xuereb}},\ }\bibfield  {title} {\bibinfo {title} {Perspective on quantum
  thermodynamics},\ }\href {https://doi.org/10.1088/1367-2630/18/1/011002}
  {\bibfield  {journal} {\bibinfo  {journal} {New Journal of Physics}\ }\textbf
  {\bibinfo {volume} {18}},\ \bibinfo {pages} {011002} (\bibinfo {year}
  {2016})}\BibitemShut {NoStop}%
\bibitem [{\citenamefont {Auff\`eves}(2022)}]{PRXQuantum.3.020101}%
  \BibitemOpen
  \bibfield  {author} {\bibinfo {author} {\bibfnamefont {A.}~\bibnamefont
  {Auff\`eves}},\ }\bibfield  {title} {\bibinfo {title} {{Quantum Technologies
  Need a Quantum Energy Initiative}},\ }\href
  {https://doi.org/10.1103/PRXQuantum.3.020101} {\bibfield  {journal} {\bibinfo
   {journal} {PRX Quantum}\ }\textbf {\bibinfo {volume} {3}},\ \bibinfo {pages}
  {020101} (\bibinfo {year} {2022})}\BibitemShut {NoStop}%
\bibitem [{\citenamefont {Hovhannisyan}\ \emph {et~al.}(2013)\citenamefont
  {Hovhannisyan}, \citenamefont {Perarnau-Llobet}, \citenamefont {Huber},\ and\
  \citenamefont {Ac\'{\i}n}}]{PhysRevLett.111.240401}%
  \BibitemOpen
  \bibfield  {author} {\bibinfo {author} {\bibfnamefont {K.~V.}\ \bibnamefont
  {Hovhannisyan}}, \bibinfo {author} {\bibfnamefont {M.}~\bibnamefont
  {Perarnau-Llobet}}, \bibinfo {author} {\bibfnamefont {M.}~\bibnamefont
  {Huber}},\ and\ \bibinfo {author} {\bibfnamefont {A.}~\bibnamefont
  {Ac\'{\i}n}},\ }\bibfield  {title} {\bibinfo {title} {{Entanglement
  Generation is Not Necessary for Optimal Work Extraction}},\ }\href
  {https://doi.org/10.1103/PhysRevLett.111.240401} {\bibfield  {journal}
  {\bibinfo  {journal} {Phys. Rev. Lett.}\ }\textbf {\bibinfo {volume} {111}},\
  \bibinfo {pages} {240401} (\bibinfo {year} {2013})}\BibitemShut {NoStop}%
\bibitem [{\citenamefont {Guryanova}\ \emph {et~al.}(2016)\citenamefont
  {Guryanova}, \citenamefont {Popescu}, \citenamefont {Short}, \citenamefont
  {Silva},\ and\ \citenamefont {Skrzypczyk}}]{Guryanova2016}%
  \BibitemOpen
  \bibfield  {author} {\bibinfo {author} {\bibfnamefont {Y.}~\bibnamefont
  {Guryanova}}, \bibinfo {author} {\bibfnamefont {S.}~\bibnamefont {Popescu}},
  \bibinfo {author} {\bibfnamefont {A.~J.}\ \bibnamefont {Short}}, \bibinfo
  {author} {\bibfnamefont {R.}~\bibnamefont {Silva}},\ and\ \bibinfo {author}
  {\bibfnamefont {P.}~\bibnamefont {Skrzypczyk}},\ }\bibfield  {title}
  {\bibinfo {title} {Thermodynamics of quantum systems with multiple conserved
  quantities},\ }\href {https://doi.org/10.1038/ncomms12049} {\bibfield
  {journal} {\bibinfo  {journal} {Nature Communications}\ }\textbf {\bibinfo
  {volume} {7}},\ \bibinfo {pages} {12049} (\bibinfo {year}
  {2016})}\BibitemShut {NoStop}%
\bibitem [{\citenamefont {Campaioli}\ \emph {et~al.}(2017)\citenamefont
  {Campaioli}, \citenamefont {Pollock}, \citenamefont {Binder}, \citenamefont
  {C\'eleri}, \citenamefont {Goold}, \citenamefont {Vinjanampathy},\ and\
  \citenamefont {Modi}}]{PhysRevLett.118.150601}%
  \BibitemOpen
  \bibfield  {author} {\bibinfo {author} {\bibfnamefont {F.}~\bibnamefont
  {Campaioli}}, \bibinfo {author} {\bibfnamefont {F.~A.}\ \bibnamefont
  {Pollock}}, \bibinfo {author} {\bibfnamefont {F.~C.}\ \bibnamefont {Binder}},
  \bibinfo {author} {\bibfnamefont {L.}~\bibnamefont {C\'eleri}}, \bibinfo
  {author} {\bibfnamefont {J.}~\bibnamefont {Goold}}, \bibinfo {author}
  {\bibfnamefont {S.}~\bibnamefont {Vinjanampathy}},\ and\ \bibinfo {author}
  {\bibfnamefont {K.}~\bibnamefont {Modi}},\ }\bibfield  {title} {\bibinfo
  {title} {{Enhancing the Charging Power of Quantum Batteries}},\ }\href
  {https://doi.org/10.1103/PhysRevLett.118.150601} {\bibfield  {journal}
  {\bibinfo  {journal} {Phys. Rev. Lett.}\ }\textbf {\bibinfo {volume} {118}},\
  \bibinfo {pages} {150601} (\bibinfo {year} {2017})}\BibitemShut {NoStop}%
\bibitem [{\citenamefont {Campaioli}\ \emph {et~al.}(2018)\citenamefont
  {Campaioli}, \citenamefont {Pollock},\ and\ \citenamefont
  {Vinjanampathy}}]{Campaioli2018}%
  \BibitemOpen
  \bibfield  {author} {\bibinfo {author} {\bibfnamefont {F.}~\bibnamefont
  {Campaioli}}, \bibinfo {author} {\bibfnamefont {F.~A.}\ \bibnamefont
  {Pollock}},\ and\ \bibinfo {author} {\bibfnamefont {S.}~\bibnamefont
  {Vinjanampathy}},\ }\bibinfo {title} {Quantum batteries},\ in\ \href
  {https://doi.org/10.1007/978-3-319-99046-0_8} {\emph {\bibinfo {booktitle}
  {Thermodynamics in the Quantum Regime: Fundamental Aspects and New
  Directions}}}\ (\bibinfo  {publisher} {Springer International Publishing},\
  \bibinfo {address} {Cham},\ \bibinfo {year} {2018})\ pp.\ \bibinfo {pages}
  {207--225}\BibitemShut {NoStop}%
\bibitem [{\citenamefont {Barra}(2019)}]{PhysRevLett.122.210601}%
  \BibitemOpen
  \bibfield  {author} {\bibinfo {author} {\bibfnamefont {F.}~\bibnamefont
  {Barra}},\ }\bibfield  {title} {\bibinfo {title} {{Dissipative Charging of a
  Quantum Battery}},\ }\href {https://doi.org/10.1103/PhysRevLett.122.210601}
  {\bibfield  {journal} {\bibinfo  {journal} {Phys. Rev. Lett.}\ }\textbf
  {\bibinfo {volume} {122}},\ \bibinfo {pages} {210601} (\bibinfo {year}
  {2019})}\BibitemShut {NoStop}%
\bibitem [{\citenamefont {Monsel}\ \emph {et~al.}(2020)\citenamefont {Monsel},
  \citenamefont {Fellous-Asiani}, \citenamefont {Huard},\ and\ \citenamefont
  {Auff\`eves}}]{PhysRevLett.124.130601}%
  \BibitemOpen
  \bibfield  {author} {\bibinfo {author} {\bibfnamefont {J.}~\bibnamefont
  {Monsel}}, \bibinfo {author} {\bibfnamefont {M.}~\bibnamefont
  {Fellous-Asiani}}, \bibinfo {author} {\bibfnamefont {B.}~\bibnamefont
  {Huard}},\ and\ \bibinfo {author} {\bibfnamefont {A.}~\bibnamefont
  {Auff\`eves}},\ }\bibfield  {title} {\bibinfo {title} {{The Energetic Cost of
  Work Extraction}},\ }\href {https://doi.org/10.1103/PhysRevLett.124.130601}
  {\bibfield  {journal} {\bibinfo  {journal} {Phys. Rev. Lett.}\ }\textbf
  {\bibinfo {volume} {124}},\ \bibinfo {pages} {130601} (\bibinfo {year}
  {2020})}\BibitemShut {NoStop}%
\bibitem [{\citenamefont {Garc\'{\i}a-Pintos}\ \emph
  {et~al.}(2020)\citenamefont {Garc\'{\i}a-Pintos}, \citenamefont {Hamma},\
  and\ \citenamefont {del Campo}}]{PhysRevLett.125.040601}%
  \BibitemOpen
  \bibfield  {author} {\bibinfo {author} {\bibfnamefont {L.~P.}\ \bibnamefont
  {Garc\'{\i}a-Pintos}}, \bibinfo {author} {\bibfnamefont {A.}~\bibnamefont
  {Hamma}},\ and\ \bibinfo {author} {\bibfnamefont {A.}~\bibnamefont {del
  Campo}},\ }\bibfield  {title} {\bibinfo {title} {{Fluctuations in Extractable
  Work Bound the Charging Power of Quantum Batteries}},\ }\href
  {https://doi.org/10.1103/PhysRevLett.125.040601} {\bibfield  {journal}
  {\bibinfo  {journal} {Phys. Rev. Lett.}\ }\textbf {\bibinfo {volume} {125}},\
  \bibinfo {pages} {040601} (\bibinfo {year} {2020})}\BibitemShut {NoStop}%
\bibitem [{\citenamefont {Rossini}\ \emph {et~al.}(2020)\citenamefont
  {Rossini}, \citenamefont {Andolina}, \citenamefont {Rosa}, \citenamefont
  {Carrega},\ and\ \citenamefont {Polini}}]{PhysRevLett.125.236402}%
  \BibitemOpen
  \bibfield  {author} {\bibinfo {author} {\bibfnamefont {D.}~\bibnamefont
  {Rossini}}, \bibinfo {author} {\bibfnamefont {G.~M.}\ \bibnamefont
  {Andolina}}, \bibinfo {author} {\bibfnamefont {D.}~\bibnamefont {Rosa}},
  \bibinfo {author} {\bibfnamefont {M.}~\bibnamefont {Carrega}},\ and\ \bibinfo
  {author} {\bibfnamefont {M.}~\bibnamefont {Polini}},\ }\bibfield  {title}
  {\bibinfo {title} {{Quantum Advantage in the Charging Process of
  Sachdev-Ye-Kitaev Batteries}},\ }\href
  {https://doi.org/10.1103/PhysRevLett.125.236402} {\bibfield  {journal}
  {\bibinfo  {journal} {Phys. Rev. Lett.}\ }\textbf {\bibinfo {volume} {125}},\
  \bibinfo {pages} {236402} (\bibinfo {year} {2020})}\BibitemShut {NoStop}%
\bibitem [{\citenamefont {Seah}\ \emph {et~al.}(2021)\citenamefont {Seah},
  \citenamefont {Perarnau-Llobet}, \citenamefont {Haack}, \citenamefont
  {Brunner},\ and\ \citenamefont {Nimmrichter}}]{PhysRevLett.127.100601}%
  \BibitemOpen
  \bibfield  {author} {\bibinfo {author} {\bibfnamefont {S.}~\bibnamefont
  {Seah}}, \bibinfo {author} {\bibfnamefont {M.}~\bibnamefont
  {Perarnau-Llobet}}, \bibinfo {author} {\bibfnamefont {G.}~\bibnamefont
  {Haack}}, \bibinfo {author} {\bibfnamefont {N.}~\bibnamefont {Brunner}},\
  and\ \bibinfo {author} {\bibfnamefont {S.}~\bibnamefont {Nimmrichter}},\
  }\bibfield  {title} {\bibinfo {title} {{Quantum Speed-Up in Collisional
  Battery Charging}},\ }\href {https://doi.org/10.1103/PhysRevLett.127.100601}
  {\bibfield  {journal} {\bibinfo  {journal} {Phys. Rev. Lett.}\ }\textbf
  {\bibinfo {volume} {127}},\ \bibinfo {pages} {100601} (\bibinfo {year}
  {2021})}\BibitemShut {NoStop}%
\bibitem [{\citenamefont {Yang}\ \emph {et~al.}(2023)\citenamefont {Yang},
  \citenamefont {Yang}, \citenamefont {Alimuddin}, \citenamefont {Salvia},
  \citenamefont {Fei}, \citenamefont {Zhao}, \citenamefont {Nimmrichter},\ and\
  \citenamefont {Luo}}]{PhysRevLett.131.030402}%
  \BibitemOpen
  \bibfield  {author} {\bibinfo {author} {\bibfnamefont {X.}~\bibnamefont
  {Yang}}, \bibinfo {author} {\bibfnamefont {Y.-H.}\ \bibnamefont {Yang}},
  \bibinfo {author} {\bibfnamefont {M.}~\bibnamefont {Alimuddin}}, \bibinfo
  {author} {\bibfnamefont {R.}~\bibnamefont {Salvia}}, \bibinfo {author}
  {\bibfnamefont {S.-M.}\ \bibnamefont {Fei}}, \bibinfo {author} {\bibfnamefont
  {L.-M.}\ \bibnamefont {Zhao}}, \bibinfo {author} {\bibfnamefont
  {S.}~\bibnamefont {Nimmrichter}},\ and\ \bibinfo {author} {\bibfnamefont
  {M.-X.}\ \bibnamefont {Luo}},\ }\bibfield  {title} {\bibinfo {title}
  {{Battery Capacity of Energy-Storing Quantum Systems}},\ }\href
  {https://doi.org/10.1103/PhysRevLett.131.030402} {\bibfield  {journal}
  {\bibinfo  {journal} {Phys. Rev. Lett.}\ }\textbf {\bibinfo {volume} {131}},\
  \bibinfo {pages} {030402} (\bibinfo {year} {2023})}\BibitemShut {NoStop}%
\bibitem [{\citenamefont {Ahmadi}\ \emph {et~al.}(2024)\citenamefont {Ahmadi},
  \citenamefont {Mazurek}, \citenamefont {Horodecki},\ and\ \citenamefont
  {Barzanjeh}}]{PhysRevLett.132.210402}%
  \BibitemOpen
  \bibfield  {author} {\bibinfo {author} {\bibfnamefont {B.}~\bibnamefont
  {Ahmadi}}, \bibinfo {author} {\bibfnamefont {P.}~\bibnamefont {Mazurek}},
  \bibinfo {author} {\bibfnamefont {P.}~\bibnamefont {Horodecki}},\ and\
  \bibinfo {author} {\bibfnamefont {S.}~\bibnamefont {Barzanjeh}},\ }\bibfield
  {title} {\bibinfo {title} {{Nonreciprocal Quantum Batteries}},\ }\href
  {https://doi.org/10.1103/PhysRevLett.132.210402} {\bibfield  {journal}
  {\bibinfo  {journal} {Phys. Rev. Lett.}\ }\textbf {\bibinfo {volume} {132}},\
  \bibinfo {pages} {210402} (\bibinfo {year} {2024})}\BibitemShut {NoStop}%
\bibitem [{\citenamefont {Bhattacharyya}\ \emph {et~al.}(2024)\citenamefont
  {Bhattacharyya}, \citenamefont {Sen},\ and\ \citenamefont
  {Sen}}]{PhysRevLett.132.240401}%
  \BibitemOpen
  \bibfield  {author} {\bibinfo {author} {\bibfnamefont {A.}~\bibnamefont
  {Bhattacharyya}}, \bibinfo {author} {\bibfnamefont {K.}~\bibnamefont {Sen}},\
  and\ \bibinfo {author} {\bibfnamefont {U.}~\bibnamefont {Sen}},\ }\bibfield
  {title} {\bibinfo {title} {{Noncompletely Positive Quantum Maps Enable
  Efficient Local Energy Extraction in Batteries}},\ }\href
  {https://doi.org/10.1103/PhysRevLett.132.240401} {\bibfield  {journal}
  {\bibinfo  {journal} {Phys. Rev. Lett.}\ }\textbf {\bibinfo {volume} {132}},\
  \bibinfo {pages} {240401} (\bibinfo {year} {2024})}\BibitemShut {NoStop}%
\bibitem [{\citenamefont {Campaioli}\ \emph {et~al.}(2024)\citenamefont
  {Campaioli}, \citenamefont {Gherardini}, \citenamefont {Quach}, \citenamefont
  {Polini},\ and\ \citenamefont {Andolina}}]{RevModPhys.96.031001}%
  \BibitemOpen
  \bibfield  {author} {\bibinfo {author} {\bibfnamefont {F.}~\bibnamefont
  {Campaioli}}, \bibinfo {author} {\bibfnamefont {S.}~\bibnamefont
  {Gherardini}}, \bibinfo {author} {\bibfnamefont {J.~Q.}\ \bibnamefont
  {Quach}}, \bibinfo {author} {\bibfnamefont {M.}~\bibnamefont {Polini}},\ and\
  \bibinfo {author} {\bibfnamefont {G.~M.}\ \bibnamefont {Andolina}},\
  }\bibfield  {title} {\bibinfo {title} {{Colloquium: Quantum batteries}},\
  }\href {https://doi.org/10.1103/RevModPhys.96.031001} {\bibfield  {journal}
  {\bibinfo  {journal} {Rev. Mod. Phys.}\ }\textbf {\bibinfo {volume} {96}},\
  \bibinfo {pages} {031001} (\bibinfo {year} {2024})}\BibitemShut {NoStop}%
\bibitem [{\citenamefont {Erdman}\ \emph {et~al.}(2024)\citenamefont {Erdman},
  \citenamefont {Andolina}, \citenamefont {Giovannetti},\ and\ \citenamefont
  {No\'e}}]{PhysRevLett.133.243602}%
  \BibitemOpen
  \bibfield  {author} {\bibinfo {author} {\bibfnamefont {P.~A.}\ \bibnamefont
  {Erdman}}, \bibinfo {author} {\bibfnamefont {G.~M.}\ \bibnamefont
  {Andolina}}, \bibinfo {author} {\bibfnamefont {V.}~\bibnamefont
  {Giovannetti}},\ and\ \bibinfo {author} {\bibfnamefont {F.}~\bibnamefont
  {No\'e}},\ }\bibfield  {title} {\bibinfo {title} {Reinforcement learning
  optimization of the charging of a dicke quantum battery},\ }\href
  {https://doi.org/10.1103/PhysRevLett.133.243602} {\bibfield  {journal}
  {\bibinfo  {journal} {Phys. Rev. Lett.}\ }\textbf {\bibinfo {volume} {133}},\
  \bibinfo {pages} {243602} (\bibinfo {year} {2024})}\BibitemShut {NoStop}%
\bibitem [{\citenamefont {Lu}\ \emph {et~al.}(2025)\citenamefont {Lu},
  \citenamefont {Tian}, \citenamefont {L\"u},\ and\ \citenamefont
  {Shang}}]{PhysRevLett.134.180401}%
  \BibitemOpen
  \bibfield  {author} {\bibinfo {author} {\bibfnamefont {Z.-G.}\ \bibnamefont
  {Lu}}, \bibinfo {author} {\bibfnamefont {G.}~\bibnamefont {Tian}}, \bibinfo
  {author} {\bibfnamefont {X.-Y.}\ \bibnamefont {L\"u}},\ and\ \bibinfo
  {author} {\bibfnamefont {C.}~\bibnamefont {Shang}},\ }\bibfield  {title}
  {\bibinfo {title} {Topological quantum batteries},\ }\href
  {https://doi.org/10.1103/PhysRevLett.134.180401} {\bibfield  {journal}
  {\bibinfo  {journal} {Phys. Rev. Lett.}\ }\textbf {\bibinfo {volume} {134}},\
  \bibinfo {pages} {180401} (\bibinfo {year} {2025})}\BibitemShut {NoStop}%
\bibitem [{\citenamefont {Horodecki}\ \emph {et~al.}(2009)\citenamefont
  {Horodecki}, \citenamefont {Horodecki}, \citenamefont {Horodecki},\ and\
  \citenamefont {Horodecki}}]{RevModPhys.81.865}%
  \BibitemOpen
  \bibfield  {author} {\bibinfo {author} {\bibfnamefont {R.}~\bibnamefont
  {Horodecki}}, \bibinfo {author} {\bibfnamefont {P.}~\bibnamefont
  {Horodecki}}, \bibinfo {author} {\bibfnamefont {M.}~\bibnamefont
  {Horodecki}},\ and\ \bibinfo {author} {\bibfnamefont {K.}~\bibnamefont
  {Horodecki}},\ }\bibfield  {title} {\bibinfo {title} {{Quantum
  entanglement}},\ }\href {https://doi.org/10.1103/RevModPhys.81.865}
  {\bibfield  {journal} {\bibinfo  {journal} {Rev. Mod. Phys.}\ }\textbf
  {\bibinfo {volume} {81}},\ \bibinfo {pages} {865} (\bibinfo {year}
  {2009})}\BibitemShut {NoStop}%
\bibitem [{\citenamefont {Streltsov}\ \emph {et~al.}(2017)\citenamefont
  {Streltsov}, \citenamefont {Adesso},\ and\ \citenamefont
  {Plenio}}]{RevModPhys.89.041003}%
  \BibitemOpen
  \bibfield  {author} {\bibinfo {author} {\bibfnamefont {A.}~\bibnamefont
  {Streltsov}}, \bibinfo {author} {\bibfnamefont {G.}~\bibnamefont {Adesso}},\
  and\ \bibinfo {author} {\bibfnamefont {M.~B.}\ \bibnamefont {Plenio}},\
  }\bibfield  {title} {\bibinfo {title} {{Colloquium: Quantum coherence as a
  resource}},\ }\href {https://doi.org/10.1103/RevModPhys.89.041003} {\bibfield
   {journal} {\bibinfo  {journal} {Rev. Mod. Phys.}\ }\textbf {\bibinfo
  {volume} {89}},\ \bibinfo {pages} {041003} (\bibinfo {year}
  {2017})}\BibitemShut {NoStop}%
\bibitem [{\citenamefont {Chitambar}\ and\ \citenamefont
  {Gour}(2019)}]{RevModPhys.91.025001}%
  \BibitemOpen
  \bibfield  {author} {\bibinfo {author} {\bibfnamefont {E.}~\bibnamefont
  {Chitambar}}\ and\ \bibinfo {author} {\bibfnamefont {G.}~\bibnamefont
  {Gour}},\ }\bibfield  {title} {\bibinfo {title} {{Quantum resource
  theories}},\ }\href {https://doi.org/10.1103/RevModPhys.91.025001} {\bibfield
   {journal} {\bibinfo  {journal} {Rev. Mod. Phys.}\ }\textbf {\bibinfo
  {volume} {91}},\ \bibinfo {pages} {025001} (\bibinfo {year}
  {2019})}\BibitemShut {NoStop}%
\bibitem [{\citenamefont {Ghosh}\ \emph {et~al.}(2020)\citenamefont {Ghosh},
  \citenamefont {Chanda},\ and\ \citenamefont {Sen(De)}}]{PhysRevA.101.032115}%
  \BibitemOpen
  \bibfield  {author} {\bibinfo {author} {\bibfnamefont {S.}~\bibnamefont
  {Ghosh}}, \bibinfo {author} {\bibfnamefont {T.}~\bibnamefont {Chanda}},\ and\
  \bibinfo {author} {\bibfnamefont {A.}~\bibnamefont {Sen(De)}},\ }\bibfield
  {title} {\bibinfo {title} {{Enhancement in the performance of a quantum
  battery by ordered and disordered interactions}},\ }\href
  {https://doi.org/10.1103/PhysRevA.101.032115} {\bibfield  {journal} {\bibinfo
   {journal} {Phys. Rev. A}\ }\textbf {\bibinfo {volume} {101}},\ \bibinfo
  {pages} {032115} (\bibinfo {year} {2020})}\BibitemShut {NoStop}%
\bibitem [{\citenamefont {Kamin}\ \emph {et~al.}(2020)\citenamefont {Kamin},
  \citenamefont {Tabesh}, \citenamefont {Salimi},\ and\ \citenamefont
  {Santos}}]{PhysRevE.102.052109}%
  \BibitemOpen
  \bibfield  {author} {\bibinfo {author} {\bibfnamefont {F.~H.}\ \bibnamefont
  {Kamin}}, \bibinfo {author} {\bibfnamefont {F.~T.}\ \bibnamefont {Tabesh}},
  \bibinfo {author} {\bibfnamefont {S.}~\bibnamefont {Salimi}},\ and\ \bibinfo
  {author} {\bibfnamefont {A.~C.}\ \bibnamefont {Santos}},\ }\bibfield  {title}
  {\bibinfo {title} {{Entanglement, coherence, and charging process of quantum
  batteries}},\ }\href {https://doi.org/10.1103/PhysRevE.102.052109} {\bibfield
   {journal} {\bibinfo  {journal} {Phys. Rev. E}\ }\textbf {\bibinfo {volume}
  {102}},\ \bibinfo {pages} {052109} (\bibinfo {year} {2020})}\BibitemShut
  {NoStop}%
\bibitem [{\citenamefont {Crescente}\ \emph {et~al.}(2020)\citenamefont
  {Crescente}, \citenamefont {Carrega}, \citenamefont {Sassetti},\ and\
  \citenamefont {Ferraro}}]{PhysRevB.102.245407}%
  \BibitemOpen
  \bibfield  {author} {\bibinfo {author} {\bibfnamefont {A.}~\bibnamefont
  {Crescente}}, \bibinfo {author} {\bibfnamefont {M.}~\bibnamefont {Carrega}},
  \bibinfo {author} {\bibfnamefont {M.}~\bibnamefont {Sassetti}},\ and\
  \bibinfo {author} {\bibfnamefont {D.}~\bibnamefont {Ferraro}},\ }\bibfield
  {title} {\bibinfo {title} {{Ultrafast charging in a two-photon Dicke quantum
  battery}},\ }\href {https://doi.org/10.1103/PhysRevB.102.245407} {\bibfield
  {journal} {\bibinfo  {journal} {Phys. Rev. B}\ }\textbf {\bibinfo {volume}
  {102}},\ \bibinfo {pages} {245407} (\bibinfo {year} {2020})}\BibitemShut
  {NoStop}%
\bibitem [{\citenamefont {Huangfu}\ and\ \citenamefont
  {Jing}(2021)}]{PhysRevE.104.024129}%
  \BibitemOpen
  \bibfield  {author} {\bibinfo {author} {\bibfnamefont {Y.}~\bibnamefont
  {Huangfu}}\ and\ \bibinfo {author} {\bibfnamefont {J.}~\bibnamefont {Jing}},\
  }\bibfield  {title} {\bibinfo {title} {{High-capacity and high-power
  collective charging with spin chargers}},\ }\href
  {https://doi.org/10.1103/PhysRevE.104.024129} {\bibfield  {journal} {\bibinfo
   {journal} {Phys. Rev. E}\ }\textbf {\bibinfo {volume} {104}},\ \bibinfo
  {pages} {024129} (\bibinfo {year} {2021})}\BibitemShut {NoStop}%
\bibitem [{\citenamefont {Ghosh}\ \emph {et~al.}(2021)\citenamefont {Ghosh},
  \citenamefont {Chanda}, \citenamefont {Mal},\ and\ \citenamefont
  {Sen(De)}}]{PhysRevA.104.032207}%
  \BibitemOpen
  \bibfield  {author} {\bibinfo {author} {\bibfnamefont {S.}~\bibnamefont
  {Ghosh}}, \bibinfo {author} {\bibfnamefont {T.}~\bibnamefont {Chanda}},
  \bibinfo {author} {\bibfnamefont {S.}~\bibnamefont {Mal}},\ and\ \bibinfo
  {author} {\bibfnamefont {A.}~\bibnamefont {Sen(De)}},\ }\bibfield  {title}
  {\bibinfo {title} {{Fast charging of a quantum battery assisted by noise}},\
  }\href {https://doi.org/10.1103/PhysRevA.104.032207} {\bibfield  {journal}
  {\bibinfo  {journal} {Phys. Rev. A}\ }\textbf {\bibinfo {volume} {104}},\
  \bibinfo {pages} {032207} (\bibinfo {year} {2021})}\BibitemShut {NoStop}%
\bibitem [{\citenamefont {Liu}\ \emph {et~al.}(2021)\citenamefont {Liu},
  \citenamefont {Shi}, \citenamefont {Shi}, \citenamefont {Wang},\ and\
  \citenamefont {Yang}}]{PhysRevB.104.245418}%
  \BibitemOpen
  \bibfield  {author} {\bibinfo {author} {\bibfnamefont {J.-X.}\ \bibnamefont
  {Liu}}, \bibinfo {author} {\bibfnamefont {H.-L.}\ \bibnamefont {Shi}},
  \bibinfo {author} {\bibfnamefont {Y.-H.}\ \bibnamefont {Shi}}, \bibinfo
  {author} {\bibfnamefont {X.-H.}\ \bibnamefont {Wang}},\ and\ \bibinfo
  {author} {\bibfnamefont {W.-L.}\ \bibnamefont {Yang}},\ }\bibfield  {title}
  {\bibinfo {title} {{Entanglement and work extraction in the central-spin
  quantum battery}},\ }\href {https://doi.org/10.1103/PhysRevB.104.245418}
  {\bibfield  {journal} {\bibinfo  {journal} {Phys. Rev. B}\ }\textbf {\bibinfo
  {volume} {104}},\ \bibinfo {pages} {245418} (\bibinfo {year}
  {2021})}\BibitemShut {NoStop}%
\bibitem [{\citenamefont {Dou}\ \emph {et~al.}(2022)\citenamefont {Dou},
  \citenamefont {Lu}, \citenamefont {Wang},\ and\ \citenamefont
  {Sun}}]{PhysRevB.105.115405}%
  \BibitemOpen
  \bibfield  {author} {\bibinfo {author} {\bibfnamefont {F.-Q.}\ \bibnamefont
  {Dou}}, \bibinfo {author} {\bibfnamefont {Y.-Q.}\ \bibnamefont {Lu}},
  \bibinfo {author} {\bibfnamefont {Y.-J.}\ \bibnamefont {Wang}},\ and\
  \bibinfo {author} {\bibfnamefont {J.-A.}\ \bibnamefont {Sun}},\ }\bibfield
  {title} {\bibinfo {title} {{Extended Dicke quantum battery with interatomic
  interactions and driving field}},\ }\href
  {https://doi.org/10.1103/PhysRevB.105.115405} {\bibfield  {journal} {\bibinfo
   {journal} {Phys. Rev. B}\ }\textbf {\bibinfo {volume} {105}},\ \bibinfo
  {pages} {115405} (\bibinfo {year} {2022})}\BibitemShut {NoStop}%
\bibitem [{\citenamefont {Arjmandi}\ \emph
  {et~al.}(2022{\natexlab{a}})\citenamefont {Arjmandi}, \citenamefont {Shokri},
  \citenamefont {Faizi},\ and\ \citenamefont
  {Mohammadi}}]{PhysRevA.106.062609}%
  \BibitemOpen
  \bibfield  {author} {\bibinfo {author} {\bibfnamefont {M.~B.}\ \bibnamefont
  {Arjmandi}}, \bibinfo {author} {\bibfnamefont {A.}~\bibnamefont {Shokri}},
  \bibinfo {author} {\bibfnamefont {E.}~\bibnamefont {Faizi}},\ and\ \bibinfo
  {author} {\bibfnamefont {H.}~\bibnamefont {Mohammadi}},\ }\bibfield  {title}
  {\bibinfo {title} {{Performance of quantum batteries with correlated and
  uncorrelated chargers}},\ }\href
  {https://doi.org/10.1103/PhysRevA.106.062609} {\bibfield  {journal} {\bibinfo
   {journal} {Phys. Rev. A}\ }\textbf {\bibinfo {volume} {106}},\ \bibinfo
  {pages} {062609} (\bibinfo {year} {2022}{\natexlab{a}})}\BibitemShut
  {NoStop}%
\bibitem [{\citenamefont {Imai}\ \emph {et~al.}(2023)\citenamefont {Imai},
  \citenamefont {G\"uhne},\ and\ \citenamefont
  {Nimmrichter}}]{PhysRevA.107.022215}%
  \BibitemOpen
  \bibfield  {author} {\bibinfo {author} {\bibfnamefont {S.}~\bibnamefont
  {Imai}}, \bibinfo {author} {\bibfnamefont {O.}~\bibnamefont {G\"uhne}},\ and\
  \bibinfo {author} {\bibfnamefont {S.}~\bibnamefont {Nimmrichter}},\
  }\bibfield  {title} {\bibinfo {title} {{Work fluctuations and entanglement in
  quantum batteries}},\ }\href {https://doi.org/10.1103/PhysRevA.107.022215}
  {\bibfield  {journal} {\bibinfo  {journal} {Phys. Rev. A}\ }\textbf {\bibinfo
  {volume} {107}},\ \bibinfo {pages} {022215} (\bibinfo {year}
  {2023})}\BibitemShut {NoStop}%
\bibitem [{\citenamefont {Centrone}\ \emph {et~al.}(2023)\citenamefont
  {Centrone}, \citenamefont {Mancino},\ and\ \citenamefont
  {Paternostro}}]{PhysRevA.108.052213}%
  \BibitemOpen
  \bibfield  {author} {\bibinfo {author} {\bibfnamefont {F.}~\bibnamefont
  {Centrone}}, \bibinfo {author} {\bibfnamefont {L.}~\bibnamefont {Mancino}},\
  and\ \bibinfo {author} {\bibfnamefont {M.}~\bibnamefont {Paternostro}},\
  }\bibfield  {title} {\bibinfo {title} {{Charging batteries with quantum
  squeezing}},\ }\href {https://doi.org/10.1103/PhysRevA.108.052213} {\bibfield
   {journal} {\bibinfo  {journal} {Phys. Rev. A}\ }\textbf {\bibinfo {volume}
  {108}},\ \bibinfo {pages} {052213} (\bibinfo {year} {2023})}\BibitemShut
  {NoStop}%
\bibitem [{\citenamefont {Ma}\ \emph {et~al.}(2024)\citenamefont {Ma},
  \citenamefont {Xu}, \citenamefont {Li}, \citenamefont {Li},\ and\
  \citenamefont {Zhu}}]{PhysRevA.110.022433}%
  \BibitemOpen
  \bibfield  {author} {\bibinfo {author} {\bibfnamefont {H.-B.}\ \bibnamefont
  {Ma}}, \bibinfo {author} {\bibfnamefont {K.}~\bibnamefont {Xu}}, \bibinfo
  {author} {\bibfnamefont {H.-G.}\ \bibnamefont {Li}}, \bibinfo {author}
  {\bibfnamefont {Z.-G.}\ \bibnamefont {Li}},\ and\ \bibinfo {author}
  {\bibfnamefont {H.-J.}\ \bibnamefont {Zhu}},\ }\bibfield  {title} {\bibinfo
  {title} {{Enhancing the charging performance of quantum batteries with the
  work medium of an entangled coupled-cavity array}},\ }\href
  {https://doi.org/10.1103/PhysRevA.110.022433} {\bibfield  {journal} {\bibinfo
   {journal} {Phys. Rev. A}\ }\textbf {\bibinfo {volume} {110}},\ \bibinfo
  {pages} {022433} (\bibinfo {year} {2024})}\BibitemShut {NoStop}%
\bibitem [{\citenamefont {Grazi}\ \emph {et~al.}(2024)\citenamefont {Grazi},
  \citenamefont {Sacco~Shaikh}, \citenamefont {Sassetti}, \citenamefont
  {Traverso~Ziani},\ and\ \citenamefont {Ferraro}}]{PhysRevLett.133.197001}%
  \BibitemOpen
  \bibfield  {author} {\bibinfo {author} {\bibfnamefont {R.}~\bibnamefont
  {Grazi}}, \bibinfo {author} {\bibfnamefont {D.}~\bibnamefont {Sacco~Shaikh}},
  \bibinfo {author} {\bibfnamefont {M.}~\bibnamefont {Sassetti}}, \bibinfo
  {author} {\bibfnamefont {N.}~\bibnamefont {Traverso~Ziani}},\ and\ \bibinfo
  {author} {\bibfnamefont {D.}~\bibnamefont {Ferraro}},\ }\bibfield  {title}
  {\bibinfo {title} {Controlling energy storage crossing quantum phase
  transitions in an integrable spin quantum battery},\ }\href
  {https://doi.org/10.1103/PhysRevLett.133.197001} {\bibfield  {journal}
  {\bibinfo  {journal} {Phys. Rev. Lett.}\ }\textbf {\bibinfo {volume} {133}},\
  \bibinfo {pages} {197001} (\bibinfo {year} {2024})}\BibitemShut {NoStop}%
\bibitem [{\citenamefont {Pokhrel}\ and\ \citenamefont
  {Gea-Banacloche}(2025)}]{PhysRevLett.134.130401}%
  \BibitemOpen
  \bibfield  {author} {\bibinfo {author} {\bibfnamefont {S.}~\bibnamefont
  {Pokhrel}}\ and\ \bibinfo {author} {\bibfnamefont {J.}~\bibnamefont
  {Gea-Banacloche}},\ }\bibfield  {title} {\bibinfo {title} {Large collective
  power enhancement in dissipative charging of a quantum battery},\ }\href
  {https://doi.org/10.1103/PhysRevLett.134.130401} {\bibfield  {journal}
  {\bibinfo  {journal} {Phys. Rev. Lett.}\ }\textbf {\bibinfo {volume} {134}},\
  \bibinfo {pages} {130401} (\bibinfo {year} {2025})}\BibitemShut {NoStop}%
\bibitem [{\citenamefont {Ferraro}\ \emph {et~al.}(2018)\citenamefont
  {Ferraro}, \citenamefont {Campisi}, \citenamefont {Andolina}, \citenamefont
  {Pellegrini},\ and\ \citenamefont {Polini}}]{PhysRevLett.120.117702}%
  \BibitemOpen
  \bibfield  {author} {\bibinfo {author} {\bibfnamefont {D.}~\bibnamefont
  {Ferraro}}, \bibinfo {author} {\bibfnamefont {M.}~\bibnamefont {Campisi}},
  \bibinfo {author} {\bibfnamefont {G.~M.}\ \bibnamefont {Andolina}}, \bibinfo
  {author} {\bibfnamefont {V.}~\bibnamefont {Pellegrini}},\ and\ \bibinfo
  {author} {\bibfnamefont {M.}~\bibnamefont {Polini}},\ }\bibfield  {title}
  {\bibinfo {title} {{High-Power Collective Charging of a Solid-State Quantum
  Battery}},\ }\href {https://doi.org/10.1103/PhysRevLett.120.117702}
  {\bibfield  {journal} {\bibinfo  {journal} {Phys. Rev. Lett.}\ }\textbf
  {\bibinfo {volume} {120}},\ \bibinfo {pages} {117702} (\bibinfo {year}
  {2018})}\BibitemShut {NoStop}%
\bibitem [{\citenamefont {Gyhm}\ \emph {et~al.}(2022)\citenamefont {Gyhm},
  \citenamefont {\ifmmode~\check{S}\else \v{S}\fi{}afr\'anek},\ and\
  \citenamefont {Rosa}}]{PhysRevLett.128.140501}%
  \BibitemOpen
  \bibfield  {author} {\bibinfo {author} {\bibfnamefont {J.-Y.}\ \bibnamefont
  {Gyhm}}, \bibinfo {author} {\bibfnamefont {D.}~\bibnamefont
  {\ifmmode~\check{S}\else \v{S}\fi{}afr\'anek}},\ and\ \bibinfo {author}
  {\bibfnamefont {D.}~\bibnamefont {Rosa}},\ }\bibfield  {title} {\bibinfo
  {title} {{Quantum Charging Advantage Cannot Be Extensive without Global
  Operations}},\ }\href {https://doi.org/10.1103/PhysRevLett.128.140501}
  {\bibfield  {journal} {\bibinfo  {journal} {Phys. Rev. Lett.}\ }\textbf
  {\bibinfo {volume} {128}},\ \bibinfo {pages} {140501} (\bibinfo {year}
  {2022})}\BibitemShut {NoStop}%
\bibitem [{\citenamefont {Alicki}\ and\ \citenamefont
  {Fannes}(2013)}]{PhysRevE.87.042123}%
  \BibitemOpen
  \bibfield  {author} {\bibinfo {author} {\bibfnamefont {R.}~\bibnamefont
  {Alicki}}\ and\ \bibinfo {author} {\bibfnamefont {M.}~\bibnamefont
  {Fannes}},\ }\bibfield  {title} {\bibinfo {title} {Entanglement boost for
  extractable work from ensembles of quantum batteries},\ }\href
  {https://doi.org/10.1103/PhysRevE.87.042123} {\bibfield  {journal} {\bibinfo
  {journal} {Phys. Rev. E}\ }\textbf {\bibinfo {volume} {87}},\ \bibinfo
  {pages} {042123} (\bibinfo {year} {2013})}\BibitemShut {NoStop}%
\bibitem [{\citenamefont {Santos}(2021)}]{PhysRevE.103.042118}%
  \BibitemOpen
  \bibfield  {author} {\bibinfo {author} {\bibfnamefont {A.~C.}\ \bibnamefont
  {Santos}},\ }\bibfield  {title} {\bibinfo {title} {Quantum advantage of
  two-level batteries in the self-discharging process},\ }\href
  {https://doi.org/10.1103/PhysRevE.103.042118} {\bibfield  {journal} {\bibinfo
   {journal} {Phys. Rev. E}\ }\textbf {\bibinfo {volume} {103}},\ \bibinfo
  {pages} {042118} (\bibinfo {year} {2021})}\BibitemShut {NoStop}%
\bibitem [{\citenamefont {Arjmandi}\ \emph
  {et~al.}(2022{\natexlab{b}})\citenamefont {Arjmandi}, \citenamefont
  {Mohammadi},\ and\ \citenamefont {Santos}}]{PhysRevE.105.054115}%
  \BibitemOpen
  \bibfield  {author} {\bibinfo {author} {\bibfnamefont {M.~B.}\ \bibnamefont
  {Arjmandi}}, \bibinfo {author} {\bibfnamefont {H.}~\bibnamefont
  {Mohammadi}},\ and\ \bibinfo {author} {\bibfnamefont {A.~C.}\ \bibnamefont
  {Santos}},\ }\bibfield  {title} {\bibinfo {title} {{Enhancing
  self-discharging process with disordered quantum batteries}},\ }\href
  {https://doi.org/10.1103/PhysRevE.105.054115} {\bibfield  {journal} {\bibinfo
   {journal} {Phys. Rev. E}\ }\textbf {\bibinfo {volume} {105}},\ \bibinfo
  {pages} {054115} (\bibinfo {year} {2022}{\natexlab{b}})}\BibitemShut
  {NoStop}%
\bibitem [{\citenamefont {Xu}\ \emph {et~al.}(2024)\citenamefont {Xu},
  \citenamefont {Li}, \citenamefont {Zhu},\ and\ \citenamefont
  {Liu}}]{PhysRevE.109.054132}%
  \BibitemOpen
  \bibfield  {author} {\bibinfo {author} {\bibfnamefont {K.}~\bibnamefont
  {Xu}}, \bibinfo {author} {\bibfnamefont {H.-G.}\ \bibnamefont {Li}}, \bibinfo
  {author} {\bibfnamefont {H.-J.}\ \bibnamefont {Zhu}},\ and\ \bibinfo {author}
  {\bibfnamefont {W.-M.}\ \bibnamefont {Liu}},\ }\bibfield  {title} {\bibinfo
  {title} {{Inhibiting the self-discharging process of quantum batteries in
  non-Markovian noises}},\ }\href {https://doi.org/10.1103/PhysRevE.109.054132}
  {\bibfield  {journal} {\bibinfo  {journal} {Phys. Rev. E}\ }\textbf {\bibinfo
  {volume} {109}},\ \bibinfo {pages} {054132} (\bibinfo {year}
  {2024})}\BibitemShut {NoStop}%
\bibitem [{\citenamefont {Bai}\ and\ \citenamefont
  {An}(2020)}]{PhysRevA.102.060201}%
  \BibitemOpen
  \bibfield  {author} {\bibinfo {author} {\bibfnamefont {S.-Y.}\ \bibnamefont
  {Bai}}\ and\ \bibinfo {author} {\bibfnamefont {J.-H.}\ \bibnamefont {An}},\
  }\bibfield  {title} {\bibinfo {title} {{Floquet engineering to reactivate a
  dissipative quantum battery}},\ }\href
  {https://doi.org/10.1103/PhysRevA.102.060201} {\bibfield  {journal} {\bibinfo
   {journal} {Phys. Rev. A}\ }\textbf {\bibinfo {volume} {102}},\ \bibinfo
  {pages} {060201(R)} (\bibinfo {year} {2020})}\BibitemShut {NoStop}%
\bibitem [{\citenamefont {Song}\ \emph {et~al.}(2024)\citenamefont {Song},
  \citenamefont {Liu}, \citenamefont {Zhou}, \citenamefont {Yang},\ and\
  \citenamefont {An}}]{PhysRevLett.132.090401}%
  \BibitemOpen
  \bibfield  {author} {\bibinfo {author} {\bibfnamefont {W.-L.}\ \bibnamefont
  {Song}}, \bibinfo {author} {\bibfnamefont {H.-B.}\ \bibnamefont {Liu}},
  \bibinfo {author} {\bibfnamefont {B.}~\bibnamefont {Zhou}}, \bibinfo {author}
  {\bibfnamefont {W.-L.}\ \bibnamefont {Yang}},\ and\ \bibinfo {author}
  {\bibfnamefont {J.-H.}\ \bibnamefont {An}},\ }\bibfield  {title} {\bibinfo
  {title} {{Remote Charging and Degradation Suppression for the Quantum
  Battery}},\ }\href {https://doi.org/10.1103/PhysRevLett.132.090401}
  {\bibfield  {journal} {\bibinfo  {journal} {Phys. Rev. Lett.}\ }\textbf
  {\bibinfo {volume} {132}},\ \bibinfo {pages} {090401} (\bibinfo {year}
  {2024})}\BibitemShut {NoStop}%
\bibitem [{\citenamefont {Yao}\ and\ \citenamefont
  {Shao}(2022)}]{PhysRevE.106.014138}%
  \BibitemOpen
  \bibfield  {author} {\bibinfo {author} {\bibfnamefont {Y.}~\bibnamefont
  {Yao}}\ and\ \bibinfo {author} {\bibfnamefont {X.~Q.}\ \bibnamefont {Shao}},\
  }\bibfield  {title} {\bibinfo {title} {{Optimal charging of open spin-chain
  quantum batteries via homodyne-based feedback control}},\ }\href
  {https://doi.org/10.1103/PhysRevE.106.014138} {\bibfield  {journal} {\bibinfo
   {journal} {Phys. Rev. E}\ }\textbf {\bibinfo {volume} {106}},\ \bibinfo
  {pages} {014138} (\bibinfo {year} {2022})}\BibitemShut {NoStop}%
\bibitem [{\citenamefont {Quach}\ and\ \citenamefont
  {Munro}(2020)}]{PhysRevApplied.14.024092}%
  \BibitemOpen
  \bibfield  {author} {\bibinfo {author} {\bibfnamefont {J.~Q.}\ \bibnamefont
  {Quach}}\ and\ \bibinfo {author} {\bibfnamefont {W.~J.}\ \bibnamefont
  {Munro}},\ }\bibfield  {title} {\bibinfo {title} {{Using Dark States to
  Charge and Stabilize Open Quantum Batteries}},\ }\href
  {https://doi.org/10.1103/PhysRevApplied.14.024092} {\bibfield  {journal}
  {\bibinfo  {journal} {Phys. Rev. Appl.}\ }\textbf {\bibinfo {volume} {14}},\
  \bibinfo {pages} {024092} (\bibinfo {year} {2020})}\BibitemShut {NoStop}%
\bibitem [{\citenamefont {Andolina}\ \emph {et~al.}(2019)\citenamefont
  {Andolina}, \citenamefont {Keck}, \citenamefont {Mari}, \citenamefont
  {Campisi}, \citenamefont {Giovannetti},\ and\ \citenamefont
  {Polini}}]{PhysRevLett.122.047702}%
  \BibitemOpen
  \bibfield  {author} {\bibinfo {author} {\bibfnamefont {G.~M.}\ \bibnamefont
  {Andolina}}, \bibinfo {author} {\bibfnamefont {M.}~\bibnamefont {Keck}},
  \bibinfo {author} {\bibfnamefont {A.}~\bibnamefont {Mari}}, \bibinfo {author}
  {\bibfnamefont {M.}~\bibnamefont {Campisi}}, \bibinfo {author} {\bibfnamefont
  {V.}~\bibnamefont {Giovannetti}},\ and\ \bibinfo {author} {\bibfnamefont
  {M.}~\bibnamefont {Polini}},\ }\bibfield  {title} {\bibinfo {title}
  {{Extractable Work, the Role of Correlations, and Asymptotic Freedom in
  Quantum Batteries}},\ }\href {https://doi.org/10.1103/PhysRevLett.122.047702}
  {\bibfield  {journal} {\bibinfo  {journal} {Phys. Rev. Lett.}\ }\textbf
  {\bibinfo {volume} {122}},\ \bibinfo {pages} {047702} (\bibinfo {year}
  {2019})}\BibitemShut {NoStop}%
\bibitem [{\citenamefont {Quach}\ \emph {et~al.}(2022)\citenamefont {Quach},
  \citenamefont {McGhee}, \citenamefont {Ganzer}, \citenamefont {Rouse},
  \citenamefont {Lovett}, \citenamefont {Gauger}, \citenamefont {Keeling},
  \citenamefont {Cerullo}, \citenamefont {Lidzey},\ and\ \citenamefont
  {Virgili}}]{doi:10.1126/sciadv.abk3160}%
  \BibitemOpen
  \bibfield  {author} {\bibinfo {author} {\bibfnamefont {J.~Q.}\ \bibnamefont
  {Quach}}, \bibinfo {author} {\bibfnamefont {K.~E.}\ \bibnamefont {McGhee}},
  \bibinfo {author} {\bibfnamefont {L.}~\bibnamefont {Ganzer}}, \bibinfo
  {author} {\bibfnamefont {D.~M.}\ \bibnamefont {Rouse}}, \bibinfo {author}
  {\bibfnamefont {B.~W.}\ \bibnamefont {Lovett}}, \bibinfo {author}
  {\bibfnamefont {E.~M.}\ \bibnamefont {Gauger}}, \bibinfo {author}
  {\bibfnamefont {J.}~\bibnamefont {Keeling}}, \bibinfo {author} {\bibfnamefont
  {G.}~\bibnamefont {Cerullo}}, \bibinfo {author} {\bibfnamefont {D.~G.}\
  \bibnamefont {Lidzey}},\ and\ \bibinfo {author} {\bibfnamefont
  {T.}~\bibnamefont {Virgili}},\ }\bibfield  {title} {\bibinfo {title}
  {{Superabsorption in an organic microcavity: Toward a quantum battery}},\
  }\href {https://doi.org/10.1126/sciadv.abk3160} {\bibfield  {journal}
  {\bibinfo  {journal} {Science Advances}\ }\textbf {\bibinfo {volume} {8}},\
  \bibinfo {pages} {eabk3160} (\bibinfo {year} {2022})}\BibitemShut {NoStop}%
\bibitem [{\citenamefont {Gemme}\ \emph {et~al.}(2022)\citenamefont {Gemme},
  \citenamefont {Grossi}, \citenamefont {Ferraro}, \citenamefont {Vallecorsa},\
  and\ \citenamefont {Sassetti}}]{batteries8050043}%
  \BibitemOpen
  \bibfield  {author} {\bibinfo {author} {\bibfnamefont {G.}~\bibnamefont
  {Gemme}}, \bibinfo {author} {\bibfnamefont {M.}~\bibnamefont {Grossi}},
  \bibinfo {author} {\bibfnamefont {D.}~\bibnamefont {Ferraro}}, \bibinfo
  {author} {\bibfnamefont {S.}~\bibnamefont {Vallecorsa}},\ and\ \bibinfo
  {author} {\bibfnamefont {M.}~\bibnamefont {Sassetti}},\ }\bibfield  {title}
  {\bibinfo {title} {{IBM Quantum Platforms: A Quantum Battery Perspective}},\
  }\href {https://www.mdpi.com/2313-0105/8/5/43} {\bibfield  {journal}
  {\bibinfo  {journal} {Batteries}\ }\textbf {\bibinfo {volume} {8}} (\bibinfo
  {year} {2022})}\BibitemShut {NoStop}%
\bibitem [{\citenamefont {Joshi}\ and\ \citenamefont
  {Mahesh}(2022)}]{PhysRevA.106.042601}%
  \BibitemOpen
  \bibfield  {author} {\bibinfo {author} {\bibfnamefont {J.}~\bibnamefont
  {Joshi}}\ and\ \bibinfo {author} {\bibfnamefont {T.~S.}\ \bibnamefont
  {Mahesh}},\ }\bibfield  {title} {\bibinfo {title} {{Experimental
  investigation of a quantum battery using star-topology NMR spin systems}},\
  }\href {https://doi.org/10.1103/PhysRevA.106.042601} {\bibfield  {journal}
  {\bibinfo  {journal} {Phys. Rev. A}\ }\textbf {\bibinfo {volume} {106}},\
  \bibinfo {pages} {042601} (\bibinfo {year} {2022})}\BibitemShut {NoStop}%
\bibitem [{\citenamefont {Zhu}\ \emph {et~al.}(2023)\citenamefont {Zhu},
  \citenamefont {Chen}, \citenamefont {Hasegawa},\ and\ \citenamefont
  {Xue}}]{PhysRevLett.131.240401}%
  \BibitemOpen
  \bibfield  {author} {\bibinfo {author} {\bibfnamefont {G.}~\bibnamefont
  {Zhu}}, \bibinfo {author} {\bibfnamefont {Y.}~\bibnamefont {Chen}}, \bibinfo
  {author} {\bibfnamefont {Y.}~\bibnamefont {Hasegawa}},\ and\ \bibinfo
  {author} {\bibfnamefont {P.}~\bibnamefont {Xue}},\ }\bibfield  {title}
  {\bibinfo {title} {{Charging Quantum Batteries via Indefinite Causal Order:
  Theory and Experiment}},\ }\href
  {https://doi.org/10.1103/PhysRevLett.131.240401} {\bibfield  {journal}
  {\bibinfo  {journal} {Phys. Rev. Lett.}\ }\textbf {\bibinfo {volume} {131}},\
  \bibinfo {pages} {240401} (\bibinfo {year} {2023})}\BibitemShut {NoStop}%
\bibitem [{\citenamefont {Maillette~de Buy~Wenniger}\ \emph
  {et~al.}(2023)\citenamefont {Maillette~de Buy~Wenniger}, \citenamefont
  {Thomas}, \citenamefont {Maffei}, \citenamefont {Wein}, \citenamefont {Pont},
  \citenamefont {Belabas}, \citenamefont {Prasad}, \citenamefont {Harouri},
  \citenamefont {Lema\^{\i}tre}, \citenamefont {Sagnes}, \citenamefont
  {Somaschi}, \citenamefont {Auff\`eves},\ and\ \citenamefont
  {Senellart}}]{PhysRevLett.131.260401}%
  \BibitemOpen
  \bibfield  {author} {\bibinfo {author} {\bibfnamefont {I.}~\bibnamefont
  {Maillette~de Buy~Wenniger}}, \bibinfo {author} {\bibfnamefont {S.~E.}\
  \bibnamefont {Thomas}}, \bibinfo {author} {\bibfnamefont {M.}~\bibnamefont
  {Maffei}}, \bibinfo {author} {\bibfnamefont {S.~C.}\ \bibnamefont {Wein}},
  \bibinfo {author} {\bibfnamefont {M.}~\bibnamefont {Pont}}, \bibinfo {author}
  {\bibfnamefont {N.}~\bibnamefont {Belabas}}, \bibinfo {author} {\bibfnamefont
  {S.}~\bibnamefont {Prasad}}, \bibinfo {author} {\bibfnamefont
  {A.}~\bibnamefont {Harouri}}, \bibinfo {author} {\bibfnamefont
  {A.}~\bibnamefont {Lema\^{\i}tre}}, \bibinfo {author} {\bibfnamefont
  {I.}~\bibnamefont {Sagnes}}, \bibinfo {author} {\bibfnamefont
  {N.}~\bibnamefont {Somaschi}}, \bibinfo {author} {\bibfnamefont
  {A.}~\bibnamefont {Auff\`eves}},\ and\ \bibinfo {author} {\bibfnamefont
  {P.}~\bibnamefont {Senellart}},\ }\bibfield  {title} {\bibinfo {title}
  {{Experimental Analysis of Energy Transfers between a Quantum Emitter and
  Light Fields}},\ }\href {https://doi.org/10.1103/PhysRevLett.131.260401}
  {\bibfield  {journal} {\bibinfo  {journal} {Phys. Rev. Lett.}\ }\textbf
  {\bibinfo {volume} {131}},\ \bibinfo {pages} {260401} (\bibinfo {year}
  {2023})}\BibitemShut {NoStop}%
\bibitem [{\citenamefont {Yu}\ \emph {et~al.}(2024)\citenamefont {Yu} \emph
  {et~al.}}]{PhysRevA.109.062614}%
  \BibitemOpen
  \bibfield  {author} {\bibinfo {author} {\bibfnamefont {J.}~\bibnamefont {Yu}}
  \emph {et~al.},\ }\bibfield  {title} {\bibinfo {title} {{Experimental
  demonstration of a Maxwell's demon quantum battery in a superconducting noisy
  intermediate-scale quantum processor}},\ }\href
  {https://doi.org/10.1103/PhysRevA.109.062614} {\bibfield  {journal} {\bibinfo
   {journal} {Phys. Rev. A}\ }\textbf {\bibinfo {volume} {109}},\ \bibinfo
  {pages} {062614} (\bibinfo {year} {2024})}\BibitemShut {NoStop}%
\bibitem [{\citenamefont {Dutt}\ \emph {et~al.}(2007)\citenamefont {Dutt},
  \citenamefont {Childress}, \citenamefont {Jiang}, \citenamefont {Togan},
  \citenamefont {Maze}, \citenamefont {Jelezko}, \citenamefont {Zibrov},
  \citenamefont {Hemmer},\ and\ \citenamefont
  {Lukin}}]{doi:10.1126/science.1139831}%
  \BibitemOpen
  \bibfield  {author} {\bibinfo {author} {\bibfnamefont {M.~V.~G.}\
  \bibnamefont {Dutt}}, \bibinfo {author} {\bibfnamefont {L.}~\bibnamefont
  {Childress}}, \bibinfo {author} {\bibfnamefont {L.}~\bibnamefont {Jiang}},
  \bibinfo {author} {\bibfnamefont {E.}~\bibnamefont {Togan}}, \bibinfo
  {author} {\bibfnamefont {J.}~\bibnamefont {Maze}}, \bibinfo {author}
  {\bibfnamefont {F.}~\bibnamefont {Jelezko}}, \bibinfo {author} {\bibfnamefont
  {A.~S.}\ \bibnamefont {Zibrov}}, \bibinfo {author} {\bibfnamefont {P.~R.}\
  \bibnamefont {Hemmer}},\ and\ \bibinfo {author} {\bibfnamefont {M.~D.}\
  \bibnamefont {Lukin}},\ }\bibfield  {title} {\bibinfo {title} {{Quantum
  Register Based on Individual Electronic and Nuclear Spin Qubits in
  Diamond}},\ }\href {https://doi.org/10.1126/science.1139831} {\bibfield
  {journal} {\bibinfo  {journal} {Science}\ }\textbf {\bibinfo {volume}
  {316}},\ \bibinfo {pages} {1312} (\bibinfo {year} {2007})}\BibitemShut
  {NoStop}%
\bibitem [{\citenamefont {Balasubramanian}\ \emph {et~al.}(2008)\citenamefont
  {Balasubramanian}, \citenamefont {Chan}, \citenamefont {Kolesov},
  \citenamefont {Al-Hmoud}, \citenamefont {Tisler}, \citenamefont {Shin},
  \citenamefont {Kim}, \citenamefont {Wojcik}, \citenamefont {Hemmer},
  \citenamefont {Krueger}, \citenamefont {Hanke}, \citenamefont
  {Leitenstorfer}, \citenamefont {Bratschitsch}, \citenamefont {Jelezko},\ and\
  \citenamefont {Wrachtrup}}]{Balasubramanian2008}%
  \BibitemOpen
  \bibfield  {author} {\bibinfo {author} {\bibfnamefont {G.}~\bibnamefont
  {Balasubramanian}}, \bibinfo {author} {\bibfnamefont {I.~Y.}\ \bibnamefont
  {Chan}}, \bibinfo {author} {\bibfnamefont {R.}~\bibnamefont {Kolesov}},
  \bibinfo {author} {\bibfnamefont {M.}~\bibnamefont {Al-Hmoud}}, \bibinfo
  {author} {\bibfnamefont {J.}~\bibnamefont {Tisler}}, \bibinfo {author}
  {\bibfnamefont {C.}~\bibnamefont {Shin}}, \bibinfo {author} {\bibfnamefont
  {C.}~\bibnamefont {Kim}}, \bibinfo {author} {\bibfnamefont {A.}~\bibnamefont
  {Wojcik}}, \bibinfo {author} {\bibfnamefont {P.~R.}\ \bibnamefont {Hemmer}},
  \bibinfo {author} {\bibfnamefont {A.}~\bibnamefont {Krueger}}, \bibinfo
  {author} {\bibfnamefont {T.}~\bibnamefont {Hanke}}, \bibinfo {author}
  {\bibfnamefont {A.}~\bibnamefont {Leitenstorfer}}, \bibinfo {author}
  {\bibfnamefont {R.}~\bibnamefont {Bratschitsch}}, \bibinfo {author}
  {\bibfnamefont {F.}~\bibnamefont {Jelezko}},\ and\ \bibinfo {author}
  {\bibfnamefont {J.}~\bibnamefont {Wrachtrup}},\ }\bibfield  {title} {\bibinfo
  {title} {{Nanoscale imaging magnetometry with diamond spins under ambient
  conditions}},\ }\href {https://doi.org/10.1038/nature07278} {\bibfield
  {journal} {\bibinfo  {journal} {Nature}\ }\textbf {\bibinfo {volume} {455}},\
  \bibinfo {pages} {648} (\bibinfo {year} {2008})}\BibitemShut {NoStop}%
\bibitem [{\citenamefont {Taylor}\ \emph {et~al.}(2008)\citenamefont {Taylor},
  \citenamefont {Cappellaro}, \citenamefont {Childress}, \citenamefont {Jiang},
  \citenamefont {Budker}, \citenamefont {Hemmer}, \citenamefont {Yacoby},
  \citenamefont {Walsworth},\ and\ \citenamefont {Lukin}}]{Taylor2008}%
  \BibitemOpen
  \bibfield  {author} {\bibinfo {author} {\bibfnamefont {J.~M.}\ \bibnamefont
  {Taylor}}, \bibinfo {author} {\bibfnamefont {P.}~\bibnamefont {Cappellaro}},
  \bibinfo {author} {\bibfnamefont {L.}~\bibnamefont {Childress}}, \bibinfo
  {author} {\bibfnamefont {L.}~\bibnamefont {Jiang}}, \bibinfo {author}
  {\bibfnamefont {D.}~\bibnamefont {Budker}}, \bibinfo {author} {\bibfnamefont
  {P.~R.}\ \bibnamefont {Hemmer}}, \bibinfo {author} {\bibfnamefont
  {A.}~\bibnamefont {Yacoby}}, \bibinfo {author} {\bibfnamefont
  {R.}~\bibnamefont {Walsworth}},\ and\ \bibinfo {author} {\bibfnamefont
  {M.~D.}\ \bibnamefont {Lukin}},\ }\bibfield  {title} {\bibinfo {title}
  {{High-sensitivity diamond magnetometer with nanoscale resolution}},\ }\href
  {https://doi.org/10.1038/nphys1075} {\bibfield  {journal} {\bibinfo
  {journal} {Nature Physics}\ }\textbf {\bibinfo {volume} {4}},\ \bibinfo
  {pages} {810} (\bibinfo {year} {2008})}\BibitemShut {NoStop}%
\bibitem [{\citenamefont {Fuchs}\ \emph {et~al.}(2009)\citenamefont {Fuchs},
  \citenamefont {Dobrovitski}, \citenamefont {Toyli}, \citenamefont
  {Heremans},\ and\ \citenamefont {Awschalom}}]{science.1181193}%
  \BibitemOpen
  \bibfield  {author} {\bibinfo {author} {\bibfnamefont {G.~D.}\ \bibnamefont
  {Fuchs}}, \bibinfo {author} {\bibfnamefont {V.~V.}\ \bibnamefont
  {Dobrovitski}}, \bibinfo {author} {\bibfnamefont {D.~M.}\ \bibnamefont
  {Toyli}}, \bibinfo {author} {\bibfnamefont {F.~J.}\ \bibnamefont
  {Heremans}},\ and\ \bibinfo {author} {\bibfnamefont {D.~D.}\ \bibnamefont
  {Awschalom}},\ }\bibfield  {title} {\bibinfo {title} {{Gigahertz Dynamics of
  a Strongly Driven Single Quantum Spin}},\ }\href
  {https://doi.org/10.1126/science.1181193} {\bibfield  {journal} {\bibinfo
  {journal} {Science}\ }\textbf {\bibinfo {volume} {326}},\ \bibinfo {pages}
  {1520} (\bibinfo {year} {2009})}\BibitemShut {NoStop}%
\bibitem [{\citenamefont {Fuchs}\ \emph {et~al.}(2011)\citenamefont {Fuchs},
  \citenamefont {Burkard}, \citenamefont {Klimov},\ and\ \citenamefont
  {Awschalom}}]{Fuchs2011}%
  \BibitemOpen
  \bibfield  {author} {\bibinfo {author} {\bibfnamefont {G.~D.}\ \bibnamefont
  {Fuchs}}, \bibinfo {author} {\bibfnamefont {G.}~\bibnamefont {Burkard}},
  \bibinfo {author} {\bibfnamefont {P.~V.}\ \bibnamefont {Klimov}},\ and\
  \bibinfo {author} {\bibfnamefont {D.~D.}\ \bibnamefont {Awschalom}},\
  }\bibfield  {title} {\bibinfo {title} {{A quantum memory intrinsic to single
  nitrogen-vacancy centres in diamond}},\ }\href
  {https://doi.org/10.1038/nphys2026} {\bibfield  {journal} {\bibinfo
  {journal} {Nature Physics}\ }\textbf {\bibinfo {volume} {7}},\ \bibinfo
  {pages} {789} (\bibinfo {year} {2011})}\BibitemShut {NoStop}%
\bibitem [{\citenamefont {Wolf}\ \emph {et~al.}(2015)\citenamefont {Wolf},
  \citenamefont {Neumann}, \citenamefont {Nakamura}, \citenamefont {Sumiya},
  \citenamefont {Ohshima}, \citenamefont {Isoya},\ and\ \citenamefont
  {Wrachtrup}}]{PhysRevX.5.041001}%
  \BibitemOpen
  \bibfield  {author} {\bibinfo {author} {\bibfnamefont {T.}~\bibnamefont
  {Wolf}}, \bibinfo {author} {\bibfnamefont {P.}~\bibnamefont {Neumann}},
  \bibinfo {author} {\bibfnamefont {K.}~\bibnamefont {Nakamura}}, \bibinfo
  {author} {\bibfnamefont {H.}~\bibnamefont {Sumiya}}, \bibinfo {author}
  {\bibfnamefont {T.}~\bibnamefont {Ohshima}}, \bibinfo {author} {\bibfnamefont
  {J.}~\bibnamefont {Isoya}},\ and\ \bibinfo {author} {\bibfnamefont
  {J.}~\bibnamefont {Wrachtrup}},\ }\bibfield  {title} {\bibinfo {title}
  {{Subpicotesla Diamond Magnetometry}},\ }\href
  {https://doi.org/10.1103/PhysRevX.5.041001} {\bibfield  {journal} {\bibinfo
  {journal} {Phys. Rev. X}\ }\textbf {\bibinfo {volume} {5}},\ \bibinfo {pages}
  {041001} (\bibinfo {year} {2015})}\BibitemShut {NoStop}%
\bibitem [{\citenamefont {Humphreys}\ \emph {et~al.}(2018)\citenamefont
  {Humphreys}, \citenamefont {Kalb}, \citenamefont {Morits}, \citenamefont
  {Schouten}, \citenamefont {Vermeulen}, \citenamefont {Twitchen},
  \citenamefont {Markham},\ and\ \citenamefont {Hanson}}]{Humphreys2018}%
  \BibitemOpen
  \bibfield  {author} {\bibinfo {author} {\bibfnamefont {P.~C.}\ \bibnamefont
  {Humphreys}}, \bibinfo {author} {\bibfnamefont {N.}~\bibnamefont {Kalb}},
  \bibinfo {author} {\bibfnamefont {J.~P.~J.}\ \bibnamefont {Morits}}, \bibinfo
  {author} {\bibfnamefont {R.~N.}\ \bibnamefont {Schouten}}, \bibinfo {author}
  {\bibfnamefont {R.~F.~L.}\ \bibnamefont {Vermeulen}}, \bibinfo {author}
  {\bibfnamefont {D.~J.}\ \bibnamefont {Twitchen}}, \bibinfo {author}
  {\bibfnamefont {M.}~\bibnamefont {Markham}},\ and\ \bibinfo {author}
  {\bibfnamefont {R.}~\bibnamefont {Hanson}},\ }\bibfield  {title} {\bibinfo
  {title} {{Deterministic delivery of remote entanglement on a quantum
  network}},\ }\href {https://doi.org/10.1038/s41586-018-0200-5} {\bibfield
  {journal} {\bibinfo  {journal} {Nature}\ }\textbf {\bibinfo {volume} {558}},\
  \bibinfo {pages} {268} (\bibinfo {year} {2018})}\BibitemShut {NoStop}%
\bibitem [{\citenamefont {Bradley}\ \emph {et~al.}(2019)\citenamefont
  {Bradley}, \citenamefont {Randall}, \citenamefont {Abobeih}, \citenamefont
  {Berrevoets}, \citenamefont {Degen}, \citenamefont {Bakker}, \citenamefont
  {Markham}, \citenamefont {Twitchen},\ and\ \citenamefont
  {Taminiau}}]{PhysRevX.9.031045}%
  \BibitemOpen
  \bibfield  {author} {\bibinfo {author} {\bibfnamefont {C.~E.}\ \bibnamefont
  {Bradley}}, \bibinfo {author} {\bibfnamefont {J.}~\bibnamefont {Randall}},
  \bibinfo {author} {\bibfnamefont {M.~H.}\ \bibnamefont {Abobeih}}, \bibinfo
  {author} {\bibfnamefont {R.~C.}\ \bibnamefont {Berrevoets}}, \bibinfo
  {author} {\bibfnamefont {M.~J.}\ \bibnamefont {Degen}}, \bibinfo {author}
  {\bibfnamefont {M.~A.}\ \bibnamefont {Bakker}}, \bibinfo {author}
  {\bibfnamefont {M.}~\bibnamefont {Markham}}, \bibinfo {author} {\bibfnamefont
  {D.~J.}\ \bibnamefont {Twitchen}},\ and\ \bibinfo {author} {\bibfnamefont
  {T.~H.}\ \bibnamefont {Taminiau}},\ }\bibfield  {title} {\bibinfo {title} {{A
  Ten-Qubit Solid-State Spin Register with Quantum Memory up to One Minute}},\
  }\href {https://doi.org/10.1103/PhysRevX.9.031045} {\bibfield  {journal}
  {\bibinfo  {journal} {Phys. Rev. X}\ }\textbf {\bibinfo {volume} {9}},\
  \bibinfo {pages} {031045} (\bibinfo {year} {2019})}\BibitemShut {NoStop}%
\bibitem [{\citenamefont {Soshenko}\ \emph {et~al.}(2021)\citenamefont
  {Soshenko}, \citenamefont {Bolshedvorskii}, \citenamefont {Rubinas},
  \citenamefont {Sorokin}, \citenamefont {Smolyaninov}, \citenamefont
  {Vorobyov},\ and\ \citenamefont {Akimov}}]{PhysRevLett.126.197702}%
  \BibitemOpen
  \bibfield  {author} {\bibinfo {author} {\bibfnamefont {V.~V.}\ \bibnamefont
  {Soshenko}}, \bibinfo {author} {\bibfnamefont {S.~V.}\ \bibnamefont
  {Bolshedvorskii}}, \bibinfo {author} {\bibfnamefont {O.}~\bibnamefont
  {Rubinas}}, \bibinfo {author} {\bibfnamefont {V.~N.}\ \bibnamefont
  {Sorokin}}, \bibinfo {author} {\bibfnamefont {A.~N.}\ \bibnamefont
  {Smolyaninov}}, \bibinfo {author} {\bibfnamefont {V.~V.}\ \bibnamefont
  {Vorobyov}},\ and\ \bibinfo {author} {\bibfnamefont {A.~V.}\ \bibnamefont
  {Akimov}},\ }\bibfield  {title} {\bibinfo {title} {{Nuclear Spin Gyroscope
  based on the Nitrogen Vacancy Center in Diamond}},\ }\href
  {https://doi.org/10.1103/PhysRevLett.126.197702} {\bibfield  {journal}
  {\bibinfo  {journal} {Phys. Rev. Lett.}\ }\textbf {\bibinfo {volume} {126}},\
  \bibinfo {pages} {197702} (\bibinfo {year} {2021})}\BibitemShut {NoStop}%
\bibitem [{\citenamefont {Barry}\ \emph {et~al.}(2020)\citenamefont {Barry},
  \citenamefont {Schloss}, \citenamefont {Bauch}, \citenamefont {Turner},
  \citenamefont {Hart}, \citenamefont {Pham},\ and\ \citenamefont
  {Walsworth}}]{RevModPhys.92.015004}%
  \BibitemOpen
  \bibfield  {author} {\bibinfo {author} {\bibfnamefont {J.~F.}\ \bibnamefont
  {Barry}}, \bibinfo {author} {\bibfnamefont {J.~M.}\ \bibnamefont {Schloss}},
  \bibinfo {author} {\bibfnamefont {E.}~\bibnamefont {Bauch}}, \bibinfo
  {author} {\bibfnamefont {M.~J.}\ \bibnamefont {Turner}}, \bibinfo {author}
  {\bibfnamefont {C.~A.}\ \bibnamefont {Hart}}, \bibinfo {author}
  {\bibfnamefont {L.~M.}\ \bibnamefont {Pham}},\ and\ \bibinfo {author}
  {\bibfnamefont {R.~L.}\ \bibnamefont {Walsworth}},\ }\bibfield  {title}
  {\bibinfo {title} {{Sensitivity optimization for NV-diamond magnetometry}},\
  }\href {https://doi.org/10.1103/RevModPhys.92.015004} {\bibfield  {journal}
  {\bibinfo  {journal} {Rev. Mod. Phys.}\ }\textbf {\bibinfo {volume} {92}},\
  \bibinfo {pages} {015004} (\bibinfo {year} {2020})}\BibitemShut {NoStop}%
\bibitem [{\citenamefont {Munuera-Javaloy}\ \emph {et~al.}(2023)\citenamefont
  {Munuera-Javaloy}, \citenamefont {Tobalina},\ and\ \citenamefont
  {Casanova}}]{PhysRevLett.130.133603}%
  \BibitemOpen
  \bibfield  {author} {\bibinfo {author} {\bibfnamefont {C.}~\bibnamefont
  {Munuera-Javaloy}}, \bibinfo {author} {\bibfnamefont {A.}~\bibnamefont
  {Tobalina}},\ and\ \bibinfo {author} {\bibfnamefont {J.}~\bibnamefont
  {Casanova}},\ }\bibfield  {title} {\bibinfo {title} {{High-Resolution NMR
  Spectroscopy at Large Fields with Nitrogen Vacancy Centers}},\ }\href
  {https://doi.org/10.1103/PhysRevLett.130.133603} {\bibfield  {journal}
  {\bibinfo  {journal} {Phys. Rev. Lett.}\ }\textbf {\bibinfo {volume} {130}},\
  \bibinfo {pages} {133603} (\bibinfo {year} {2023})}\BibitemShut {NoStop}%
\bibitem [{\citenamefont {Neumann}\ \emph {et~al.}(2010)\citenamefont
  {Neumann}, \citenamefont {Kolesov}, \citenamefont {Naydenov}, \citenamefont
  {Beck}, \citenamefont {Rempp}, \citenamefont {Steiner}, \citenamefont
  {Jacques}, \citenamefont {Balasubramanian}, \citenamefont {Markham},
  \citenamefont {Twitchen}, \citenamefont {Pezzagna}, \citenamefont {Meijer},
  \citenamefont {Twamley}, \citenamefont {Jelezko},\ and\ \citenamefont
  {Wrachtrup}}]{Neumann2010}%
  \BibitemOpen
  \bibfield  {author} {\bibinfo {author} {\bibfnamefont {P.}~\bibnamefont
  {Neumann}}, \bibinfo {author} {\bibfnamefont {R.}~\bibnamefont {Kolesov}},
  \bibinfo {author} {\bibfnamefont {B.}~\bibnamefont {Naydenov}}, \bibinfo
  {author} {\bibfnamefont {J.}~\bibnamefont {Beck}}, \bibinfo {author}
  {\bibfnamefont {F.}~\bibnamefont {Rempp}}, \bibinfo {author} {\bibfnamefont
  {M.}~\bibnamefont {Steiner}}, \bibinfo {author} {\bibfnamefont
  {V.}~\bibnamefont {Jacques}}, \bibinfo {author} {\bibfnamefont
  {G.}~\bibnamefont {Balasubramanian}}, \bibinfo {author} {\bibfnamefont
  {M.~L.}\ \bibnamefont {Markham}}, \bibinfo {author} {\bibfnamefont {D.~J.}\
  \bibnamefont {Twitchen}}, \bibinfo {author} {\bibfnamefont {S.}~\bibnamefont
  {Pezzagna}}, \bibinfo {author} {\bibfnamefont {J.}~\bibnamefont {Meijer}},
  \bibinfo {author} {\bibfnamefont {J.}~\bibnamefont {Twamley}}, \bibinfo
  {author} {\bibfnamefont {F.}~\bibnamefont {Jelezko}},\ and\ \bibinfo {author}
  {\bibfnamefont {J.}~\bibnamefont {Wrachtrup}},\ }\bibfield  {title} {\bibinfo
  {title} {{Quantum register based on coupled electron spins in a
  room-temperature solid}},\ }\href {https://doi.org/10.1038/nphys1536}
  {\bibfield  {journal} {\bibinfo  {journal} {Nature Physics}\ }\textbf
  {\bibinfo {volume} {6}},\ \bibinfo {pages} {249} (\bibinfo {year}
  {2010})}\BibitemShut {NoStop}%
\bibitem [{\citenamefont {Haase}\ \emph {et~al.}(2018)\citenamefont {Haase},
  \citenamefont {Vetter}, \citenamefont {Unden}, \citenamefont {Smirne},
  \citenamefont {Rosskopf}, \citenamefont {Naydenov}, \citenamefont {Stacey},
  \citenamefont {Jelezko}, \citenamefont {Plenio},\ and\ \citenamefont
  {Huelga}}]{PhysRevLett.121.060401}%
  \BibitemOpen
  \bibfield  {author} {\bibinfo {author} {\bibfnamefont {J.~F.}\ \bibnamefont
  {Haase}}, \bibinfo {author} {\bibfnamefont {P.~J.}\ \bibnamefont {Vetter}},
  \bibinfo {author} {\bibfnamefont {T.}~\bibnamefont {Unden}}, \bibinfo
  {author} {\bibfnamefont {A.}~\bibnamefont {Smirne}}, \bibinfo {author}
  {\bibfnamefont {J.}~\bibnamefont {Rosskopf}}, \bibinfo {author}
  {\bibfnamefont {B.}~\bibnamefont {Naydenov}}, \bibinfo {author}
  {\bibfnamefont {A.}~\bibnamefont {Stacey}}, \bibinfo {author} {\bibfnamefont
  {F.}~\bibnamefont {Jelezko}}, \bibinfo {author} {\bibfnamefont {M.~B.}\
  \bibnamefont {Plenio}},\ and\ \bibinfo {author} {\bibfnamefont {S.~F.}\
  \bibnamefont {Huelga}},\ }\bibfield  {title} {\bibinfo {title} {{Controllable
  Non-Markovianity for a Spin Qubit in Diamond}},\ }\href
  {https://doi.org/10.1103/PhysRevLett.121.060401} {\bibfield  {journal}
  {\bibinfo  {journal} {Phys. Rev. Lett.}\ }\textbf {\bibinfo {volume} {121}},\
  \bibinfo {pages} {060401} (\bibinfo {year} {2018})}\BibitemShut {NoStop}%
\bibitem [{\citenamefont {Yang}\ \emph {et~al.}(2019)\citenamefont {Yang},
  \citenamefont {Song}, \citenamefont {An}, \citenamefont {Feng}, \citenamefont
  {Suter},\ and\ \citenamefont {Du}}]{Yang_2019}%
  \BibitemOpen
  \bibfield  {author} {\bibinfo {author} {\bibfnamefont {W.~L.}\ \bibnamefont
  {Yang}}, \bibinfo {author} {\bibfnamefont {W.~L.}\ \bibnamefont {Song}},
  \bibinfo {author} {\bibfnamefont {J.-H.}\ \bibnamefont {An}}, \bibinfo
  {author} {\bibfnamefont {M.}~\bibnamefont {Feng}}, \bibinfo {author}
  {\bibfnamefont {D.}~\bibnamefont {Suter}},\ and\ \bibinfo {author}
  {\bibfnamefont {J.}~\bibnamefont {Du}},\ }\bibfield  {title} {\bibinfo
  {title} {{Floquet engineering to entanglement protection of distant nitrogen
  vacancy centers}},\ }\href {https://doi.org/10.1088/1367-2630/aaf8f4}
  {\bibfield  {journal} {\bibinfo  {journal} {New Journal of Physics}\ }\textbf
  {\bibinfo {volume} {21}},\ \bibinfo {pages} {013007} (\bibinfo {year}
  {2019})}\BibitemShut {NoStop}%
\bibitem [{\citenamefont {Lu}\ \emph {et~al.}(2020)\citenamefont {Lu},
  \citenamefont {Zhang}, \citenamefont {Liu}, \citenamefont {Nori},
  \citenamefont {Fan},\ and\ \citenamefont {Pan}}]{PhysRevLett.124.210502}%
  \BibitemOpen
  \bibfield  {author} {\bibinfo {author} {\bibfnamefont {Y.-N.}\ \bibnamefont
  {Lu}}, \bibinfo {author} {\bibfnamefont {Y.-R.}\ \bibnamefont {Zhang}},
  \bibinfo {author} {\bibfnamefont {G.-Q.}\ \bibnamefont {Liu}}, \bibinfo
  {author} {\bibfnamefont {F.}~\bibnamefont {Nori}}, \bibinfo {author}
  {\bibfnamefont {H.}~\bibnamefont {Fan}},\ and\ \bibinfo {author}
  {\bibfnamefont {X.-Y.}\ \bibnamefont {Pan}},\ }\bibfield  {title} {\bibinfo
  {title} {{Observing Information Backflow from Controllable Non-Markovian
  Multichannels in Diamond}},\ }\href
  {https://doi.org/10.1103/PhysRevLett.124.210502} {\bibfield  {journal}
  {\bibinfo  {journal} {Phys. Rev. Lett.}\ }\textbf {\bibinfo {volume} {124}},\
  \bibinfo {pages} {210502} (\bibinfo {year} {2020})}\BibitemShut {NoStop}%
\bibitem [{\citenamefont {Doherty}\ \emph {et~al.}(2013)\citenamefont
  {Doherty}, \citenamefont {Manson}, \citenamefont {Delaney}, \citenamefont
  {Jelezko}, \citenamefont {Wrachtrup},\ and\ \citenamefont
  {Hollenberg}}]{DOHERTY20131}%
  \BibitemOpen
  \bibfield  {author} {\bibinfo {author} {\bibfnamefont {M.~W.}\ \bibnamefont
  {Doherty}}, \bibinfo {author} {\bibfnamefont {N.~B.}\ \bibnamefont {Manson}},
  \bibinfo {author} {\bibfnamefont {P.}~\bibnamefont {Delaney}}, \bibinfo
  {author} {\bibfnamefont {F.}~\bibnamefont {Jelezko}}, \bibinfo {author}
  {\bibfnamefont {J.}~\bibnamefont {Wrachtrup}},\ and\ \bibinfo {author}
  {\bibfnamefont {L.~C.}\ \bibnamefont {Hollenberg}},\ }\bibfield  {title}
  {\bibinfo {title} {{The nitrogen-vacancy colour centre in diamond}},\ }\href
  {https://doi.org/https://doi.org/10.1016/j.physrep.2013.02.001} {\bibfield
  {journal} {\bibinfo  {journal} {Physics Reports}\ }\textbf {\bibinfo {volume}
  {528}},\ \bibinfo {pages} {1} (\bibinfo {year} {2013})}\BibitemShut {NoStop}%
\bibitem [{\citenamefont {Balasubramanian}\ \emph {et~al.}(2009)\citenamefont
  {Balasubramanian}, \citenamefont {Neumann}, \citenamefont {Twitchen},
  \citenamefont {Markham}, \citenamefont {Kolesov}, \citenamefont {Mizuochi},
  \citenamefont {Isoya}, \citenamefont {Achard}, \citenamefont {Beck},
  \citenamefont {Tissler}, \citenamefont {Jacques}, \citenamefont {Hemmer},
  \citenamefont {Jelezko},\ and\ \citenamefont
  {Wrachtrup}}]{Balasubramanian2009}%
  \BibitemOpen
  \bibfield  {author} {\bibinfo {author} {\bibfnamefont {G.}~\bibnamefont
  {Balasubramanian}}, \bibinfo {author} {\bibfnamefont {P.}~\bibnamefont
  {Neumann}}, \bibinfo {author} {\bibfnamefont {D.}~\bibnamefont {Twitchen}},
  \bibinfo {author} {\bibfnamefont {M.}~\bibnamefont {Markham}}, \bibinfo
  {author} {\bibfnamefont {R.}~\bibnamefont {Kolesov}}, \bibinfo {author}
  {\bibfnamefont {N.}~\bibnamefont {Mizuochi}}, \bibinfo {author}
  {\bibfnamefont {J.}~\bibnamefont {Isoya}}, \bibinfo {author} {\bibfnamefont
  {J.}~\bibnamefont {Achard}}, \bibinfo {author} {\bibfnamefont
  {J.}~\bibnamefont {Beck}}, \bibinfo {author} {\bibfnamefont {J.}~\bibnamefont
  {Tissler}}, \bibinfo {author} {\bibfnamefont {V.}~\bibnamefont {Jacques}},
  \bibinfo {author} {\bibfnamefont {P.~R.}\ \bibnamefont {Hemmer}}, \bibinfo
  {author} {\bibfnamefont {F.}~\bibnamefont {Jelezko}},\ and\ \bibinfo {author}
  {\bibfnamefont {J.}~\bibnamefont {Wrachtrup}},\ }\bibfield  {title} {\bibinfo
  {title} {{Ultralong spin coherence time in isotopically engineered
  diamond}},\ }\href {https://doi.org/10.1038/nmat2420} {\bibfield  {journal}
  {\bibinfo  {journal} {Nature Materials}\ }\textbf {\bibinfo {volume} {8}},\
  \bibinfo {pages} {383} (\bibinfo {year} {2009})}\BibitemShut {NoStop}%
\bibitem [{\citenamefont {Allahverdyan}\ \emph {et~al.}(2004)\citenamefont
  {Allahverdyan}, \citenamefont {Balian},\ and\ \citenamefont
  {Nieuwenhuizen}}]{A.E.Allahverdyan_2004}%
  \BibitemOpen
  \bibfield  {author} {\bibinfo {author} {\bibfnamefont {A.~E.}\ \bibnamefont
  {Allahverdyan}}, \bibinfo {author} {\bibfnamefont {R.}~\bibnamefont
  {Balian}},\ and\ \bibinfo {author} {\bibfnamefont {T.~M.}\ \bibnamefont
  {Nieuwenhuizen}},\ }\bibfield  {title} {\bibinfo {title} {{Maximal work
  extraction from finite quantum systems}},\ }\href
  {https://doi.org/10.1209/epl/i2004-10101-2} {\bibfield  {journal} {\bibinfo
  {journal} {Europhysics Letters}\ }\textbf {\bibinfo {volume} {67}},\ \bibinfo
  {pages} {565} (\bibinfo {year} {2004})}\BibitemShut {NoStop}%
\bibitem [{\citenamefont {Francica}\ \emph {et~al.}(2020)\citenamefont
  {Francica}, \citenamefont {Binder}, \citenamefont {Guarnieri}, \citenamefont
  {Mitchison}, \citenamefont {Goold},\ and\ \citenamefont
  {Plastina}}]{PhysRevLett.125.180603}%
  \BibitemOpen
  \bibfield  {author} {\bibinfo {author} {\bibfnamefont {G.}~\bibnamefont
  {Francica}}, \bibinfo {author} {\bibfnamefont {F.~C.}\ \bibnamefont
  {Binder}}, \bibinfo {author} {\bibfnamefont {G.}~\bibnamefont {Guarnieri}},
  \bibinfo {author} {\bibfnamefont {M.~T.}\ \bibnamefont {Mitchison}}, \bibinfo
  {author} {\bibfnamefont {J.}~\bibnamefont {Goold}},\ and\ \bibinfo {author}
  {\bibfnamefont {F.}~\bibnamefont {Plastina}},\ }\bibfield  {title} {\bibinfo
  {title} {{Quantum Coherence and Ergotropy}},\ }\href
  {https://doi.org/10.1103/PhysRevLett.125.180603} {\bibfield  {journal}
  {\bibinfo  {journal} {Phys. Rev. Lett.}\ }\textbf {\bibinfo {volume} {125}},\
  \bibinfo {pages} {180603} (\bibinfo {year} {2020})}\BibitemShut {NoStop}%
\bibitem [{\citenamefont {Tirone}\ \emph {et~al.}(2023)\citenamefont {Tirone},
  \citenamefont {Salvia}, \citenamefont {Chessa},\ and\ \citenamefont
  {Giovannetti}}]{PhysRevLett.131.060402}%
  \BibitemOpen
  \bibfield  {author} {\bibinfo {author} {\bibfnamefont {S.}~\bibnamefont
  {Tirone}}, \bibinfo {author} {\bibfnamefont {R.}~\bibnamefont {Salvia}},
  \bibinfo {author} {\bibfnamefont {S.}~\bibnamefont {Chessa}},\ and\ \bibinfo
  {author} {\bibfnamefont {V.}~\bibnamefont {Giovannetti}},\ }\bibfield
  {title} {\bibinfo {title} {Work extraction processes from noisy quantum
  batteries: The role of nonlocal resources},\ }\href
  {https://doi.org/10.1103/PhysRevLett.131.060402} {\bibfield  {journal}
  {\bibinfo  {journal} {Phys. Rev. Lett.}\ }\textbf {\bibinfo {volume} {131}},\
  \bibinfo {pages} {060402} (\bibinfo {year} {2023})}\BibitemShut {NoStop}%
\bibitem [{\citenamefont {Baumgratz}\ \emph {et~al.}(2014)\citenamefont
  {Baumgratz}, \citenamefont {Cramer},\ and\ \citenamefont
  {Plenio}}]{PhysRevLett.113.140401}%
  \BibitemOpen
  \bibfield  {author} {\bibinfo {author} {\bibfnamefont {T.}~\bibnamefont
  {Baumgratz}}, \bibinfo {author} {\bibfnamefont {M.}~\bibnamefont {Cramer}},\
  and\ \bibinfo {author} {\bibfnamefont {M.~B.}\ \bibnamefont {Plenio}},\
  }\bibfield  {title} {\bibinfo {title} {{Quantifying Coherence}},\ }\href
  {https://doi.org/10.1103/PhysRevLett.113.140401} {\bibfield  {journal}
  {\bibinfo  {journal} {Phys. Rev. Lett.}\ }\textbf {\bibinfo {volume} {113}},\
  \bibinfo {pages} {140401} (\bibinfo {year} {2014})}\BibitemShut {NoStop}%
\bibitem [{\citenamefont {Shi}\ \emph {et~al.}(2022)\citenamefont {Shi},
  \citenamefont {Ding}, \citenamefont {Wan}, \citenamefont {Wang},\ and\
  \citenamefont {Yang}}]{PhysRevLett.129.130602}%
  \BibitemOpen
  \bibfield  {author} {\bibinfo {author} {\bibfnamefont {H.-L.}\ \bibnamefont
  {Shi}}, \bibinfo {author} {\bibfnamefont {S.}~\bibnamefont {Ding}}, \bibinfo
  {author} {\bibfnamefont {Q.-K.}\ \bibnamefont {Wan}}, \bibinfo {author}
  {\bibfnamefont {X.-H.}\ \bibnamefont {Wang}},\ and\ \bibinfo {author}
  {\bibfnamefont {W.-L.}\ \bibnamefont {Yang}},\ }\bibfield  {title} {\bibinfo
  {title} {{Entanglement, Coherence, and Extractable Work in Quantum
  Batteries}},\ }\href {https://doi.org/10.1103/PhysRevLett.129.130602}
  {\bibfield  {journal} {\bibinfo  {journal} {Phys. Rev. Lett.}\ }\textbf
  {\bibinfo {volume} {129}},\ \bibinfo {pages} {130602} (\bibinfo {year}
  {2022})}\BibitemShut {NoStop}%
\bibitem [{\citenamefont {Niu}\ \emph {et~al.}(2024)\citenamefont {Niu},
  \citenamefont {Wu}, \citenamefont {Wang}, \citenamefont {Rong},\ and\
  \citenamefont {Du}}]{PhysRevLett.133.180401}%
  \BibitemOpen
  \bibfield  {author} {\bibinfo {author} {\bibfnamefont {Z.}~\bibnamefont
  {Niu}}, \bibinfo {author} {\bibfnamefont {Y.}~\bibnamefont {Wu}}, \bibinfo
  {author} {\bibfnamefont {Y.}~\bibnamefont {Wang}}, \bibinfo {author}
  {\bibfnamefont {X.}~\bibnamefont {Rong}},\ and\ \bibinfo {author}
  {\bibfnamefont {J.}~\bibnamefont {Du}},\ }\bibfield  {title} {\bibinfo
  {title} {{Experimental Investigation of Coherent Ergotropy in a Single Spin
  System}},\ }\href {https://doi.org/10.1103/PhysRevLett.133.180401} {\bibfield
   {journal} {\bibinfo  {journal} {Phys. Rev. Lett.}\ }\textbf {\bibinfo
  {volume} {133}},\ \bibinfo {pages} {180401} (\bibinfo {year}
  {2024})}\BibitemShut {NoStop}%
\end{thebibliography}%
\end{document}